\documentclass[12pt]{article}
\usepackage[dvips]{graphicx}
\usepackage{pdproc,amsmath,amssymb,graphics,subfigure,color}

\newcommand{\de}[2]{(\delta_{#1}^d)_{#2}}

  \makeatletter 
  \def\@cite#1{[#1]} 
  \makeatother    
\begin{document}

\renewcommand{\thefootnote}{\alph{footnote}}

\title{ Supersymmetry, flavor physics and dark matter 
 %      \\  within SUSY models 
}

\author{PYUNGWON KO}

\address{ 
Department of Physics, KAIST \\
Daejon 305-701, Korea
%%%%% You may comment out the e-mail address line below.  
\\ {\rm E-mail: pko@muon.kaist.ac.kr}}

\abstract{
In this talk, 
I first discuss $b\rightarrow d$, $b\rightarrow s$ and $s\rightarrow d$ 
transitions  including time-dependent CP asymmetries in 
$B_d \rightarrow \phi K_s$ and Re ($\epsilon^{'}/\epsilon_K$). 
Then I show that  the decay $B_s \rightarrow \mu^+ \mu^-$ is useful 
to distinguish various  SUSY breaking mechanisms. 
I will also describe some possible connections between B physics and 
cosmology: 
(i) B physics and electroweak baryogenesis within SUSY models, and 
(ii) the correlation between the neutralino dark matter scattering and 
$B ( B_s \rightarrow \mu^+ \mu^- )$.  In particular, we point out that
the current upper bound on $B ( B_s \rightarrow \mu^+ \mu^- )$ form CDF 
and D0 collaborations can constrain the spin-independent 
neutralino scattering cross section more strongly than the CDMS bound.
}

\normalsize\baselineskip=15pt

\section{Introduction}

In the Standard Model (SM), flavor mixing and  CP violation in the quark 
sector have the common origin, namely the CKM mixing matrix. 
This is dictated by local gauge invariance and renormalizability 
of the SM with 3 families. 
This paradigm is well tested by many different observables in
K and B meson systems. All the data (except for possible anomalies in 
the time dependent CP asymmetries in $B_d \rightarrow \phi K_s$ and 
$B_d \rightarrow \eta^{'} K_s$  decays,  
and the baryon number asymmetry in the universe) can be accommodated 
by the CKM picture, and we have consistent understanding of flavor mixing 
and CP violation within the SM.  Despite this great success of SM, 
there are many reasons why we consider the SM merely as 
a low energy effective theory of some fundamental theory.  
In particular, quadratic divergence in the SM Higgs mass  seems to call 
for new physics beyond the SM around $\sim O(1)$ TeV. SUSY models with 
$R-$parity conservation are well motivated new physics scenarios due to 
gauge coupling unification and the presence of dark matter candidates. 
In  SUSY models, the flavor and CP structures of the soft SUSY breaking 
terms have rich structures, and 
there could be  large deviations in some processes involving 
B and K mesons, without any conflict with the current status of CKM 
phenomenology. 

In this talk, I will give a few such examples, in which we can have large 
deviations from the SM predictions, even if the CKM triangle in the SUSY 
models has the same shape as in the SM. More specifically, we 
will discuss the branching ratio  
of $B\rightarrow X_d \gamma$  and CP asymmetry therein, CP asymmetries
in $B\rightarrow X_s \gamma$ and $B_d \rightarrow \phi K_s$, 
$B_s - \overline{B_s}$ mixing (both the modulus and the phase), and 
$B_s \rightarrow \mu^+ \mu^-$ as well as $\epsilon^{'} /\epsilon_K$.  
The future  experiments at B factories 
should study these processes in greater detail, thus testing the CKM 
paradigm within the SM and exploring the flavor and CP structures of SUSY 
models.  

In phenomenological study of SUSY models, it is crucial to include  
the soft SUSY breaking terms.  However,  we do not understand the nature 
of SUSY breaking in our world, and thus we do not know the flavor and CP  
structures of soft SUSY breaking terms.  This makes it difficult to study 
flavor physics and CP violation within SUSY models, and most results are 
admittedly model dependent. 
In the following, we take two different approaches: 
(i) we use the mass insertion approximation (MIA) assuming gluino-squark 
loop contributions are dominant,  or 
(ii) we work in specific SUSY breaking scenarios which are theoretically 
well motivated.
Even if our current strategies are not perfect, our analysis method could
be used in other cases, and  we don't expect that we lose  generic 
features by such strategies. 
Eventually we will want to measure all the soft SUSY breaking 
parameters. It would not be easy to get informations on flavor and CP 
violating soft terms from LHC/NLC alone, and the low energy processes 
involving K, B mesons and $\mu, \tau$ leptons will give invaluable 
informations on flavor and/or CP violating soft SUSY breaking parameters, 
when combined with the informations on the SUSY particle mass spectra 
and flavor diagonal couplings measured at LHC and NLC. 

The plan of my talk is the following.  In Section 2 and Section 3, 
I will discuss $b\rightarrow d$ and $b\rightarrow s$  transitions 
within MIA, including $B\rightarrow X_d \gamma$  and CP asymmetry 
therein, CP asymmetries in $B\rightarrow X_s \gamma$ and 
$B_d \rightarrow \phi K_S$, and $B_s - \overline{B_s}$ mixing.
In Section 4, I discuss some correlation between $S_{\phi K}$ and 
$\epsilon^{'}/\epsilon_K$ within the $RR$ dominance scenario: namely 
the $s_R \rightarrow d_L$ transition induced by the 
$b_R \rightarrow s_R$  transition.  
In Section 5, I'll discuss the $B_s \rightarrow \mu^+ \mu^-$ as a useful 
probe of SUSY breaking mechanisms. In Section 6,  I will discuss possible 
interplay between B physics  and cosmology with two examples: 
(i) B physics and electroweak baryogenesis (EWBGEN) within SUSY models, 
and (ii) the correlation between $B_s \rightarrow \mu^+ \mu^-$ and 
the neutralino dark matter (DM) scattering  cross section. Then I conclude
in Section 7.
 
%%%%%%%%%%%%%%%%%%%%%%%%%%%%%%%%%%%%%%%%%%%%%%%%%%%%%%%%%%%%%%%%%%%%%%%%%
\section{$b\rightarrow d$ transition: $B_d - \overline{B_d}$ mixing and
CP asymmetry in $B\rightarrow X_d \gamma$ } 

In general SUSY models, squark mass matrices are not diagonal in the basis
where quark masses are diagonal. Therefore the $\tilde{g}-q_i-\tilde{q}_j$ 
vertex can change the (s)quark flavor,  leading to dangerous flavor 
changing neutral current (FCNC) processes at one loop level with strong 
interaction strength. 
Various low energy data  such as $K^0 - \overline{K^0}$ and
$B_{d(s)}^0 - \overline{B_{d(s)}^0}$ mixings, 
Re $( \epsilon^{'}/\epsilon_K )$ and $B\rightarrow X_{d(s)} \gamma$ etc. 
will put strong constraints on such flavor changing 
$\tilde{g}-q_i-\tilde{q}_j$  vertex.
In the limit of degenerate squark masses, FCNC amplitude vanishes. 
Therefore the almost degenrate squark masses may be the good starting 
point to study gluino-mediated FCNC within general SUSY models, and the 
so-called masss insertion approximation (MIA) is convenient in this case
\cite{Hall:1985dx,Gabbiani:1996hi}. In this section, we consider 
$b\rightarrow d$ transition due to gluino mediation within MIA, relegating 
the $b\rightarrow s$ and $s\rightarrow d$ transitions to the following 
sections. 

Observations of large CP violation in $B\rightarrow J/\psi K_S$
at B factories \cite{Eidelman:2004wy} 
\begin{equation}
\sin 2 \beta_{\psi K}  = ( 0.731 \pm 0.056)
\end{equation}
confirm the SM prediction, and begin to put a strong constraint on new physics
contributions to $B^0 - \overline{B^0}$ mixing and $B\rightarrow J/\psi K_S$,
when combined with 
\[
\Delta m_{B_d} = (0.502 \pm 0.007 )~{\rm ps}^{-1},
~~{\rm Br} ( B \rightarrow X_d \gamma ) <  1 \times 10^{-5}. 
\]
Here  the $B_d \rightarrow X_d \gamma$ branching ratio constraint was 
extracted from the recent experimental upper limit on the 
$B\rightarrow \rho\gamma$ branching ratio \cite{b2rho}
$B( B \rightarrow \rho \gamma ) < 2.3 \times 10^{-6}$. 
Since the decay $B\rightarrow J/\psi K_S$ is dominated by the
tree level SM process $b\rightarrow c \bar{c} s$, we expect the new physics
contribution may affect significantly the $B^0 - \overline{B^0}$ mixing
only and not the decay $B\rightarrow J/\psi K_S$. 
However, in the presence of new physics contributions to
$B^0 - \overline{B^0}$ mixing, the same new physics will generically affect
the $B\rightarrow X_d \gamma$ process \cite{kkp}, 
which is also loop suppressed in the SM \cite{ali}. 
In the following, we consider 
$B^0 - \overline{B^0}$ mixing, $B\rightarrow J/\psi K_S$ and
$B_d \rightarrow X_d \gamma$ assuming that the main SUSY contribution is 
from gluino-squark loops in addition to the usual SM contribution.

\begin{figure*}
\centering
\subfigure[$LL$ mixing]{\raisebox{1mm}
{\includegraphics[width=5.0cm,height=5.0cm]{%../susyb/bbbar/
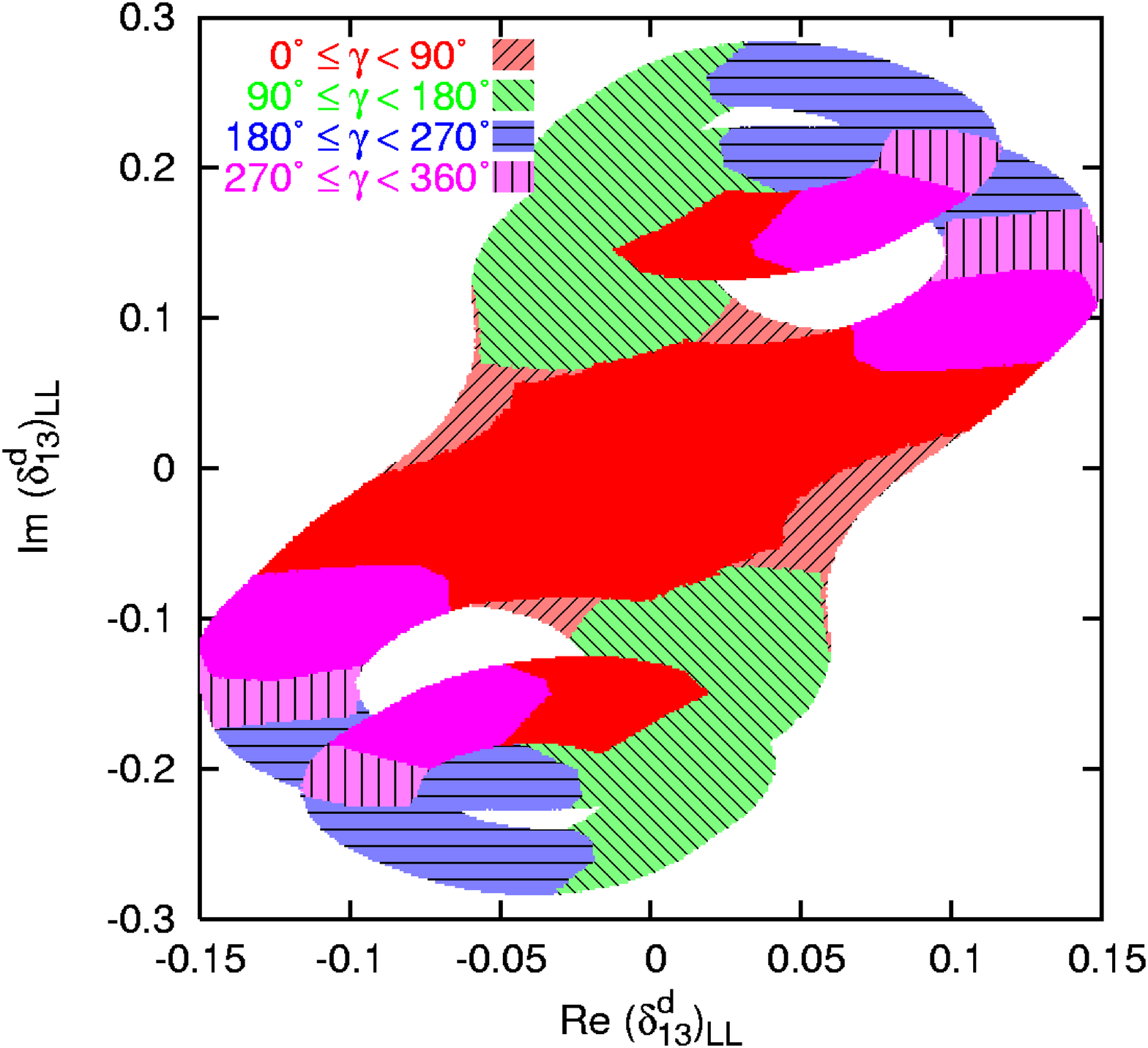}}}
\subfigure[$A_{ll}$]
{\includegraphics[width=4.8cm,height=5.0cm]
{%../susyb/bbbar/
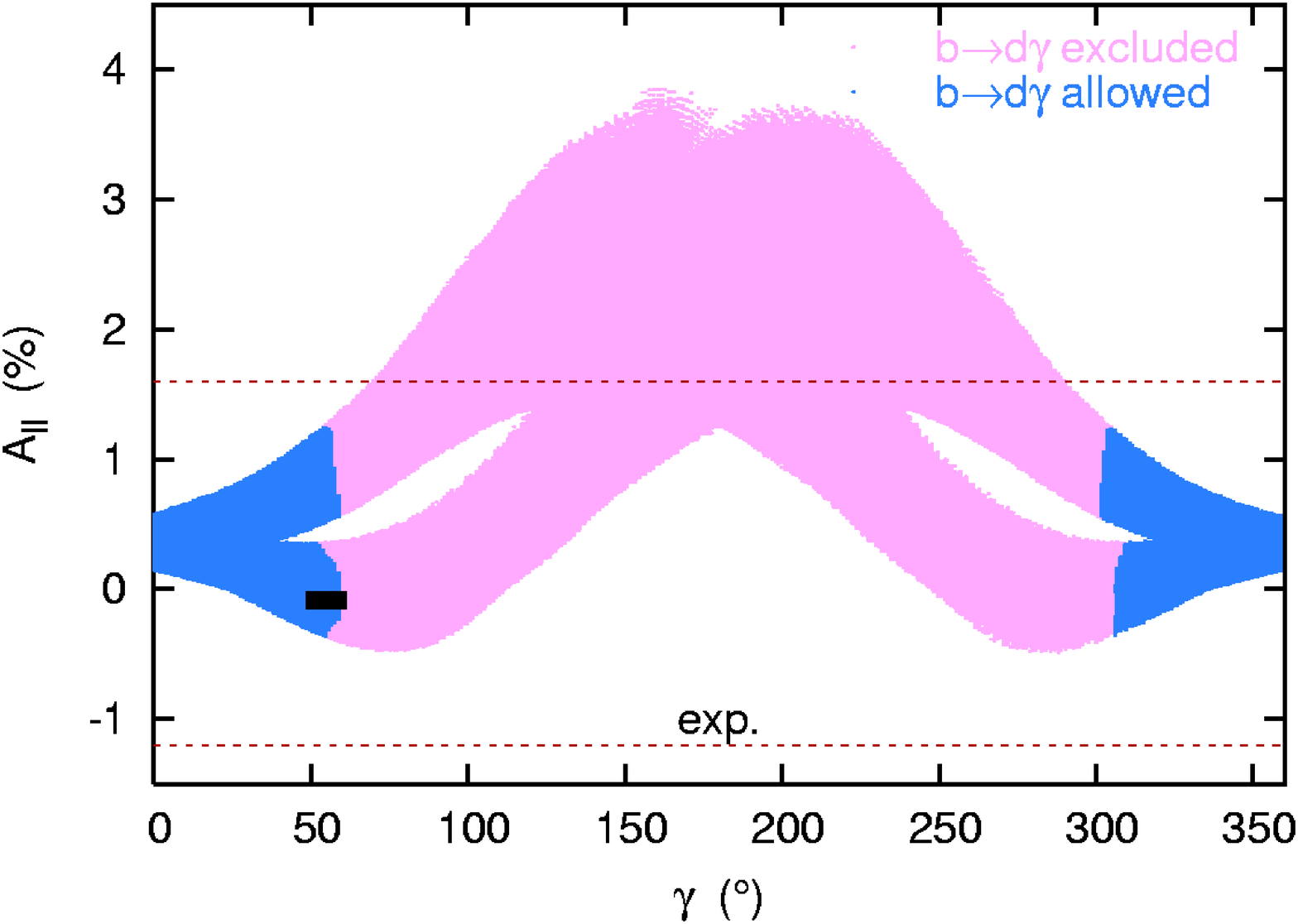}}
\subfigure[$A_{\rm CP}^{b\rightarrow d\gamma}$]
{\includegraphics[width=5.0cm,height=5.0cm]
{%../susyb/bbbar/
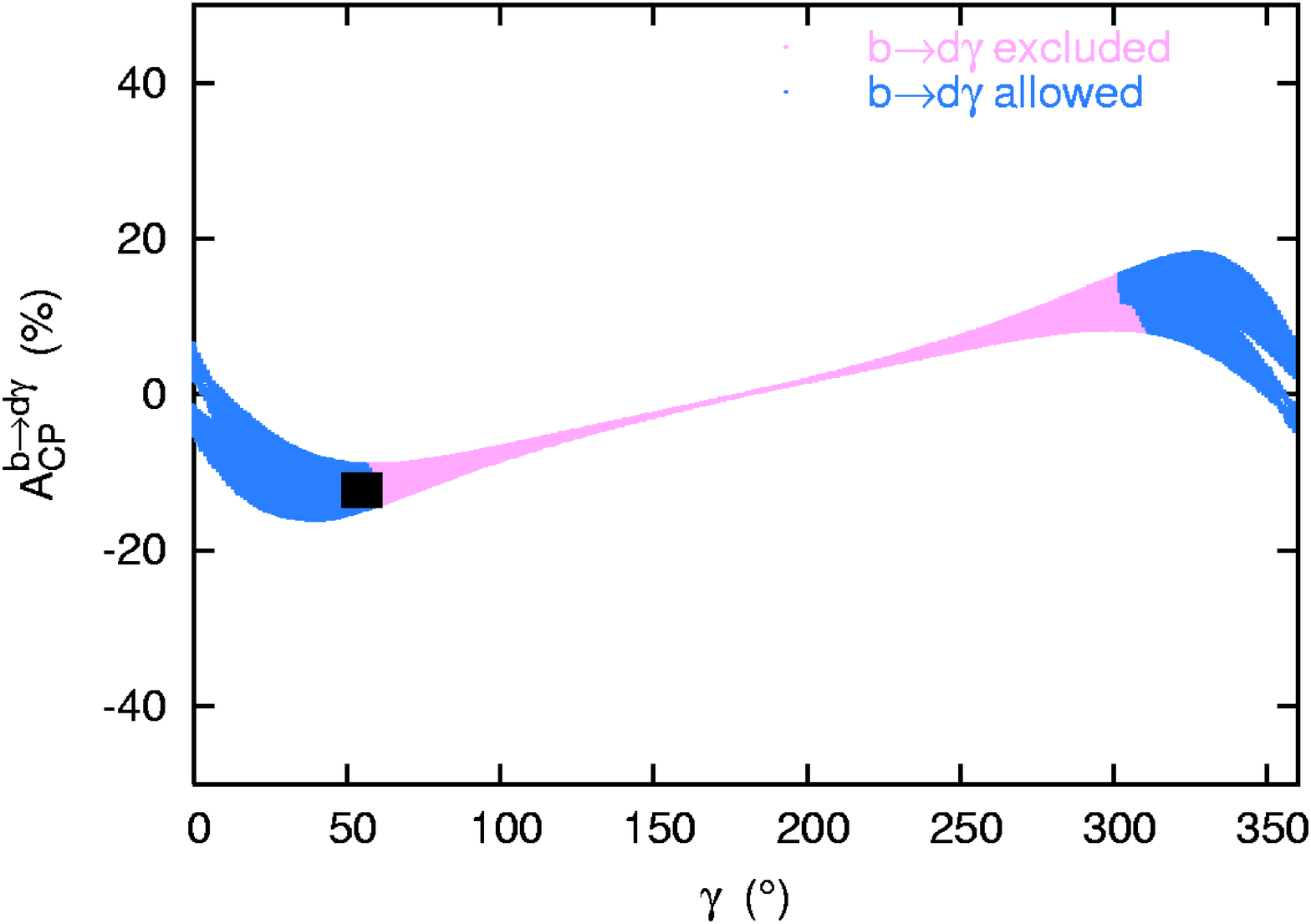}}%
\caption{
(a) The allowed range in the $LL$ insertion case for
the parameters
$( {\rm Re}(\delta_{13}^d )_{AB}, {\rm Im} (\delta_{13}^d )_{AB})$ for
different values of the KM angle $\gamma$ with different color codes:
dark (red) for $0^{\circ} \leq \gamma \leq 90^{\circ}$, light gray (green)
for $90^{\circ} \leq \gamma \leq 180^{\circ}$, very dark (blue) for
$180^{\circ} \leq \gamma \leq 270^{\circ}$ and gray (magenta) for
$270^{\circ} \leq \gamma \leq 360^{\circ}$.
The region leading to a too large branching ratio for
$B_d \rightarrow X_d \gamma$ is colored lightly and covered by parallel lines.
(b) and (c) are the dilepton charge asymmetry $A_{ll}$,  and the direct 
CP asymmetry in $B\rightarrow X_d \gamma$ as functions of $\gamma$. The 
SM predictions for $\gamma = 55^{\circ}$ are indicated by the black boxes. 
}
\label{fig:d13}
\end{figure*}%

In Fig.~\ref{fig:d13} (a), we show the allowed parameter space in
the $( {\rm Re}(\delta_{13}^d )_{LL}, {\rm Im} (\delta_{13}^d )_{LL})$ plane
for different values of the KM angle $\gamma$ with different color codes.
The region leading to too large a branching ratio for 
$B_d \rightarrow X_d \gamma$ is covered by parallel lines. Note that 
$B\rightarrow X_d \gamma$ plays an important role here.  
And the region where the dilepton charge asymmetry $A_{ll}$ 
(see Ref.~\cite{kkp} for the definition) falls out of the data 
$A_{ll}^{\rm exp} = ( 0.2 \pm 1.4 ) $ \% \cite{nir} 
within $1\sigma$ range is already excluded by the 
$B\rightarrow X_d \gamma$ branching ratio constraint [ Fig.~1 (b) ]. 
Note that the KM angle $\gamma$ should be in the range 
between $\sim - 60^\circ$ and $\sim + 60^{\circ}$, and $A_{ll}$ can have 
the opposite sign compared to the SM prediction, even if the KM angle is 
the same as its SM value $\gamma_{\rm SM} \simeq 55^{\circ}$ due to the
SUSY contributions to $B^0 - \overline{B^0}$ mixing.
In Fig.~\ref{fig:d13} (c),  we show 
the direct CP asymmetry in $B_d \rightarrow X_d \gamma$ 
as a function of the KM angles $\gamma$ for the $LL$ insertion case. 
The direct CP asymmetry is predicted to be between $\sim - 15\%$ and
$\sim +20\%$.
In the $LL$ mixing case, the SM gives the dominant contribution to
$B_d \rightarrow X_d \gamma$, but the KM angle can be different from the
SM case, because SUSY contributions to the $B^0 - \overline{B^0}$ mixing can
be significant so that the preferred value of $\gamma$ can change from 
the SM KM fitting. 
( This is the same in the rare kaon decays and the results
obtained in Ref.~\cite{ko1,ko2} apply without modifications. ) 
Therefore, it is possible to have  large deviations in the
$B_d \rightarrow X_d \gamma$ branching ratio and the direct CP violation
thereof.

\begin{figure*}[htbp]
\centering
\subfigure[$LR$ mixing]{\raisebox{1mm}
{\includegraphics[width=5.0cm,height=5.0cm]
{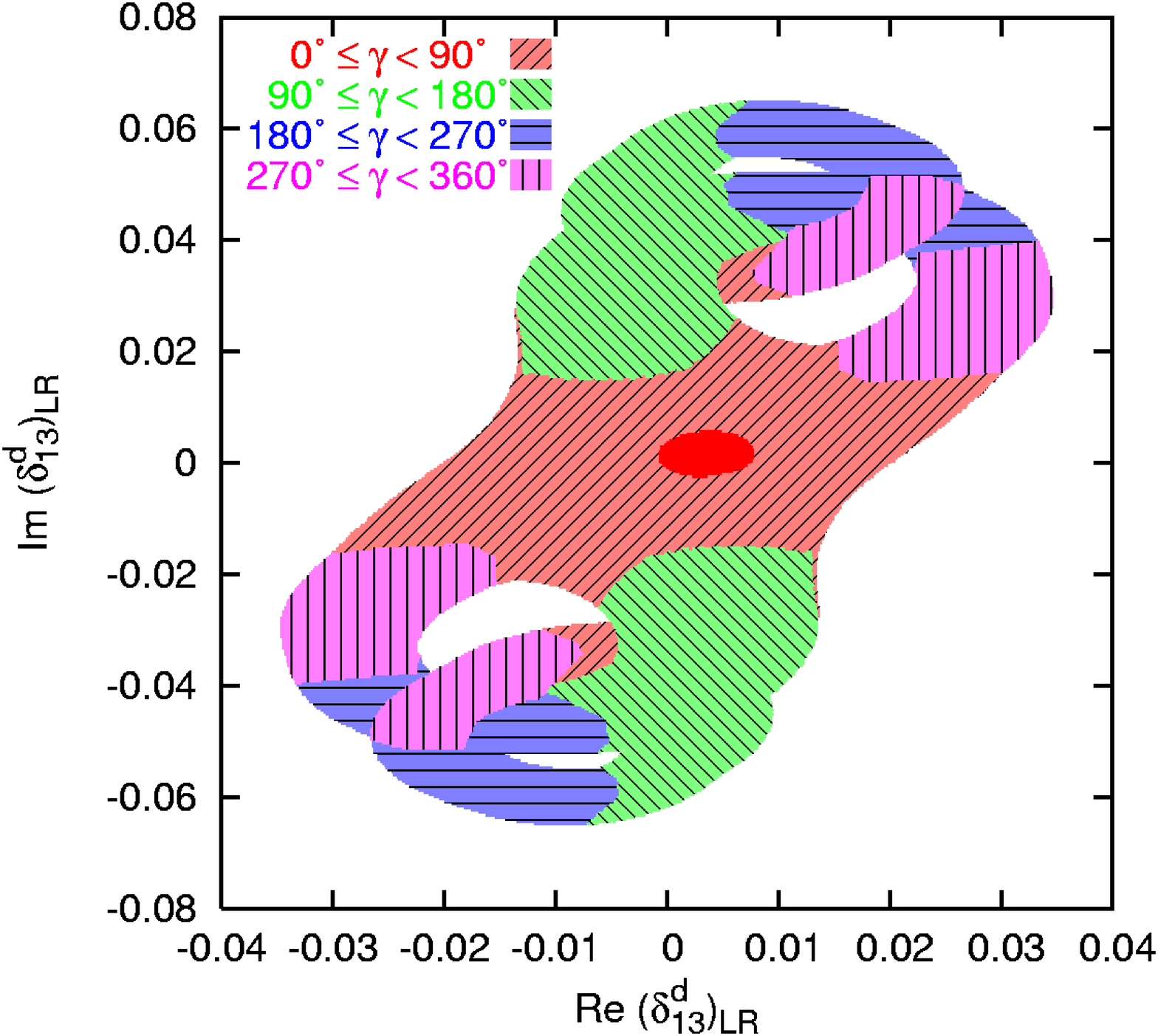}}}
\subfigure[$A_{ll}$]
{\includegraphics[width=4.8cm,height=5.0cm]
{%../susyb/bbbar/
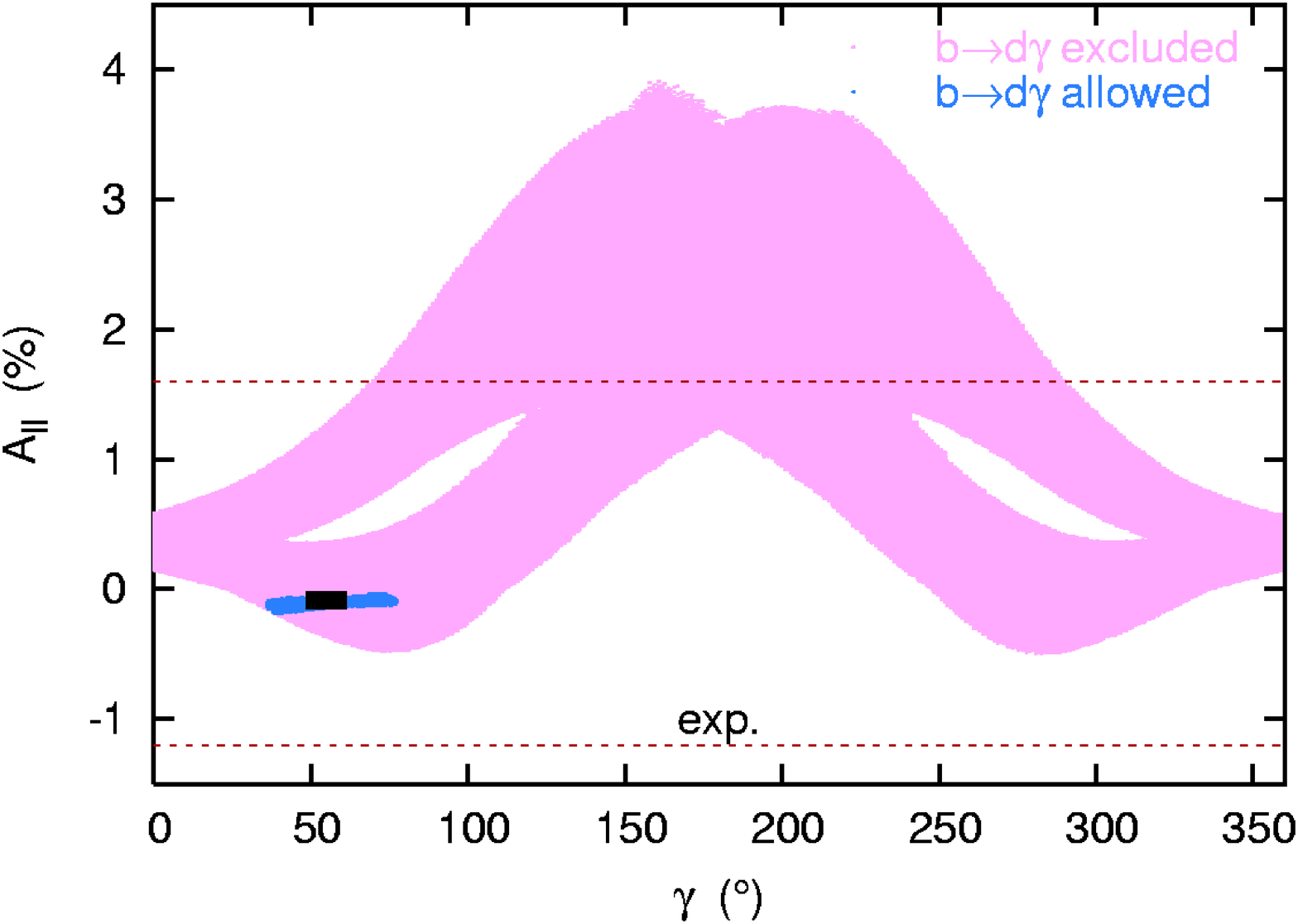}} 
\subfigure[$A_{\rm CP}^{b\rightarrow d\gamma}$]
{\includegraphics[width=5.0cm,height=5.0cm]
{%../susyb/bbbar/
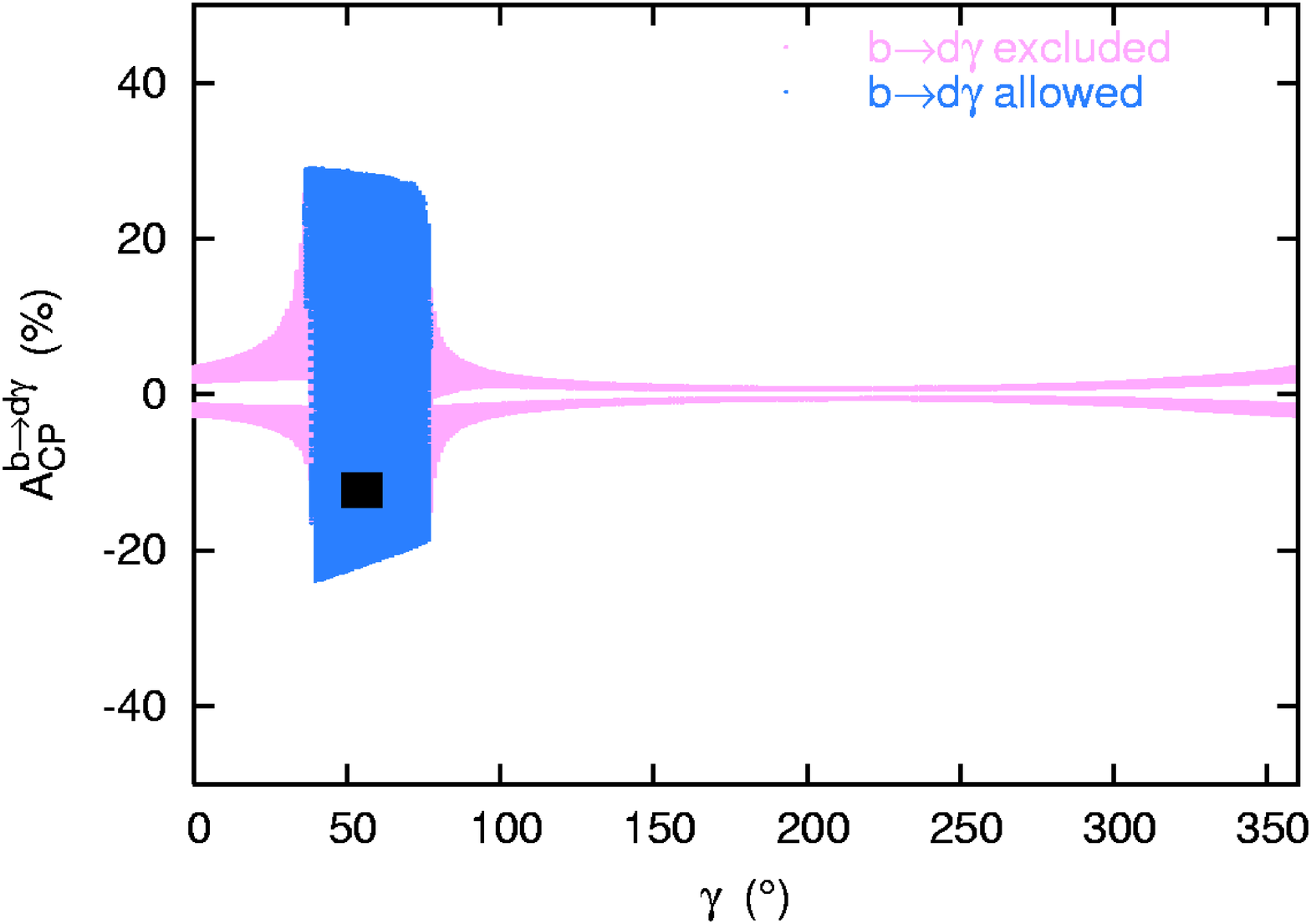}}
\caption{The $LR$ mixing case. The captions are the same as Fig~1.}
\label{fig:all}
\end{figure*}

For the $LR$ mixing, the $B(B_d \rightarrow X_d \gamma)$ puts an even 
stronger constraint  compared to the $LL$ insertion case [ Fig.~2 (a) ],  
whereas the $A_{ll}$ does not put any new constraint [ Fig.~2 (b) ].
In particular, the KM angle $\gamma$ can not be too much different from
the SM value in the $LR$ mixing case, once the
$B(B_d \rightarrow X_d \gamma)$ constraint is included.
Only $30^{\circ} \lesssim \gamma \lesssim 80^{\circ}$ is compatible with all
the data from the $B$ system, even if we do not consider the $\epsilon_K$
constraint. The resulting parameter space is significantly reduced compared
to the $LL$ insertion case. 
In Fig.~2 (b), we show the predictions for $A_{ll}$ as a function of the
KM angle $\gamma$ for the $LR$ insertion only. 
Note that the $B\rightarrow X_d \gamma$ constraint rules out
almost all the parameter space region, and the resulting $A_{ll}$ is 
essentially the same as for the SM case. In Fig.~2 (c),  we find that
there could be substantial deviation in the CP asymmetry in 
$B_d \rightarrow X_d \gamma$ from the SM predictions, 
even if the $\Delta m_B$ and $\sin 2 \beta$ is
the same as the SM predictions as well as the data. For the $LL$ insertion,
such a large deviation is possible, since the KM angle $\gamma$ can be
substantially different from the SM value. On the other hand, for the $LR$
mixing, the large deviation comes from the complex $( \delta_{13}^d )_{LR}$
even if the KM angle is set to the same value as in the SM.
The size of $( \delta_{13}^d )_{LR}$ is too small to affect the
$B^0 - \overline{B^0}$ mixing, but is still large enough to affect
$B\rightarrow X_d \gamma$. Our model independent study indicates that
the current data on the $\Delta m_B$, $\sin2\beta$ and $A_{ll}$ do still
allow a possibility for large deviations in $B\rightarrow X_d \gamma$,
both in the branching ratio and the direct CP asymmetry thereof.
These observables are indispensable to test the KM paradigm for CP 
violation completely and get ideas on possible new physics with 
new flavor/CP violation in $b\rightarrow d$ transition.

Summarizing this section, we considered the gluino-mediated SUSY 
contributions to $B^0 - \overline{B^0}$ mixing, $B\rightarrow J/\psi K_S$ 
and  $B\rightarrow X_d \gamma$ in the mass insertion approximation.
We find that the $LL$ mixing parameter can be as large as
$| (\delta_{13}^d)_{LL} | \lesssim 2 \times 10^{-1}$, but the
$LR$ mixing is strongly constrained by the $B\rightarrow X_d \gamma$
branching ratio: $| (\delta_{13}^d)_{LR} | \lesssim 10^{-2}$.
The implications for the direct CP asymmetry in $B\rightarrow X_d \gamma$
are also discussed, where substantial deviations from the SM predictions are
still possible both in the $LL$ and $LR$ insertion cases even if $\gamma 
\simeq \gamma_{\rm SM}$.
Our analysis demonstrates that all the observables, $A_{ll}$, the branching
ratio of $B\rightarrow X_d \gamma$ and the direct CP violation thereof
are very important, since they could provide informations on new flavor
and CP violation from $(\delta_{13}^d )_{LL,LR}$ (or any other new physics
scenarios with new flavor/CP violations). 
These will provide strong constraints on SUSY flavor models that attempt 
to solve hierarchies in the Yukawa  couplings and SUSY flavor problems 
using some flavor symmetry groups \cite{align}. 
Also they are indispensable in order that we can ultimately test
the KM paradigm for CP violation in the SM, since one can have very different
branching ratio and CP asymmetry for $B\rightarrow X_d \gamma$ for the SM
values of the CKM matrix elements, if there is a new physics beyond the SM
with new sources of flavor and CP violations.

%%%%%%%%%%%%%%%%%%%%%%%%%%%%%%%%%%%%%%%%%%%%%%%%%%%%%%%%%%%%%%%%%%%%%%%%%%
\section{$b\rightarrow s$ transition: $B_d \rightarrow \phi K_S$ and 
$B_s - \overline{B_s}$ mixing}

$B\rightarrow \phi K$ is a powerful testing ground for new physics. 
Because it is loop suppressed in the standard model (SM), 
this decay is very sensitive to
possible new physics contributions to $b\rightarrow s s \bar{s}$, 
a feature not shared by other charmless $B$ decays. 
Within the SM, it is dominated by the  QCD penguin diagrams 
with a top quark in the loop. 
Therefore the time dependent CP asymmetries are essentially 
the same as those in $B\rightarrow J/\psi K_S$: $\sin 2 \beta_{\phi K} 
\simeq \sin 2 \beta_{\psi K} + O(\lambda^2)$ \cite{bsss}. 

Recently both BaBar and Belle reported the branching ratio and CP asymmetries
in  the  $B_d \rightarrow \phi K_S$ decay:
\begin{eqnarray}
{\cal A}_{\phi K} (t) & \equiv &
{{\Gamma (\overline{B}^0_{\rm phys} (t) \rightarrow \phi K_S ) - 
 \Gamma (      B^0_{\rm phys} (t) \rightarrow \phi K_S )   } \over
{\Gamma (\overline{B}^0_{\rm phys} (t) \rightarrow \phi K_S ) + 
 \Gamma (      B^0_{\rm phys} (t) \rightarrow \phi K_S )   }}
\nonumber  \\
& = & - C_{\phi K} \cos ( \Delta m_B t )
  + S_{\phi K} \sin ( \Delta m_B t ),
\end{eqnarray}
where $C_{\phi K}$ and $S_{\phi K}$ are given by
\begin{equation}
C_{\phi K}  = 
{ 1 - | \lambda_{\phi K} |^2 \over 1 + | \lambda_{\phi K} |^2 } , ~~~~~
{\rm and}~~~~~
S_{\phi K}  =  
{ 2~ {\rm Im} \lambda_{\phi K} \over 1 + | \lambda_{\phi K} |^2 } ,
\end{equation}
with 
\begin{equation}
\lambda_{\phi K} \equiv - e^{ - 2 i (\beta + \theta_d )} 
{\overline{A} ( \overline{B}^0 \rightarrow \phi K_S ) \over 
A ( B^0 \rightarrow \phi K_S ) } ,
\end{equation}
and the angle $\theta_d$ represents any new physics contributions to the 
$B_d - \overline{B_d}$ mixing angle. 
The current world average is \cite{lp04} 
\[ 
\sin 2 \beta_{\phi K} = S_{\phi K} = ( 0.34 \pm 0.20 ) ,
\]
which is about 2 $\sigma$ lower than 
the SM prediction: $\sin 2 \beta_{J/\psi K_S} = ( 0.731 \pm 0.056 )$.
The direct CP asymmetry in $B_d \rightarrow \phi K_S$ is also measured,
and is  consistent with zero \cite{acp}: 
$C_{\phi K_S} = ( -0.04 \pm 0.17 )$.  

In the following, we assume that the $\tilde{b}_A - \tilde{s}_B$ (with 
$A,B = L$ or $R$) mixing has a new CP violating phase,
and study  its effects on $S_{\phi K}$,  $B\rightarrow X_s \gamma$,  
the direct CP asymmetry therein and  $B_s^0 - \overline{B_s^0}$ mixing.  
Higgs-mediated $b \rightarrow s s \bar{s}$ transition could be enhanced
for large $\tan\beta$. However, once the existing CDF limit on 
$B ( B_s \rightarrow \mu^+ \mu^-) < 5.8 \times 10^{-7}$ \cite{cdf} 
is imposed on the Higgs mediated $b\rightarrow s s \bar{s}$, 
it is found too small an effect on $S_{\phi K}$ \cite{kkkpww1,kkkpww2}. 
(The discussions on chargino loop contributions can be found in 
Ref.~\cite{khalil}.) 

We calculate the Wilson coefficients of the operators for $\Delta B=1$ 
effective Hamiltonian at the scale $\mu\sim \tilde{m}\sim m_W$. 
Then we evolve the Wilson coefficients to $\mu \sim m_b$ 
using the appropriate renormalization group (RG)  equations,  
and calculate the amplitude for $B\rightarrow \phi K$ using the 
BBNS approach~\cite{bbns}  for estimating the hadronic matrix elements.
The details of the effective Hamiltonian and the Wilson coefficients
can be found in Ref.~\cite{kkkpww1,kkkpww2}.

In the numerical analysis presented here, we 
fix the SUSY parameters to be $m_{\tilde{g}} = \tilde{m} = 400$ GeV.
In each of the mass insertion scenarios to be discussed, we vary
the mass insertions over the range $\left|\delta^d_{AB}\right|\leq
1$ to fully map the parameter space.
We then impose two important experimental constraints.
First, we demand that the predicted branching ratio for inclusive
$B\to X_s\gamma$ fall within the range
$2.0 \times 10^{-4} < B ( B\rightarrow X_s \gamma ) < 4.5 \times 10^{-4}$,
which is rather generous 
in order to  allow for various theoretical uncertainties. 
Second, we impose the current lower limit on
$\Delta M_s > 14.9$ ps$^{-1}$. 

A new CP-violating phase in $( \delta^d_{AB} )_{23}$ will also 
generate CP violation in $B\rightarrow X_s \gamma$. The current 
world average of the direct CP asymmetry 
$A_{\rm CP}^{b\rightarrow s \gamma}$ is \cite{acp} 
$A_{\rm CP}^{b\rightarrow s \gamma} = (0.5 \pm 3.6) \%$, which is now 
quite constraining (see also the discussion in Sec.~5.1 and Fig.~6).  
Within the SM, the predicted CP asymmetry is less than $\sim 0.5 \%$, and
a larger asymmetry would be a clear indication of new physics \cite{kn}.   
Where  relevant, we will show our predictions for 
$A_{\rm CP}^{b\rightarrow s \gamma}$.

\begin{figure}
{\includegraphics[width=5.0cm]
%{../../susyb/b2phik/400-1_fin/kolda/deltaLL.eps}}
{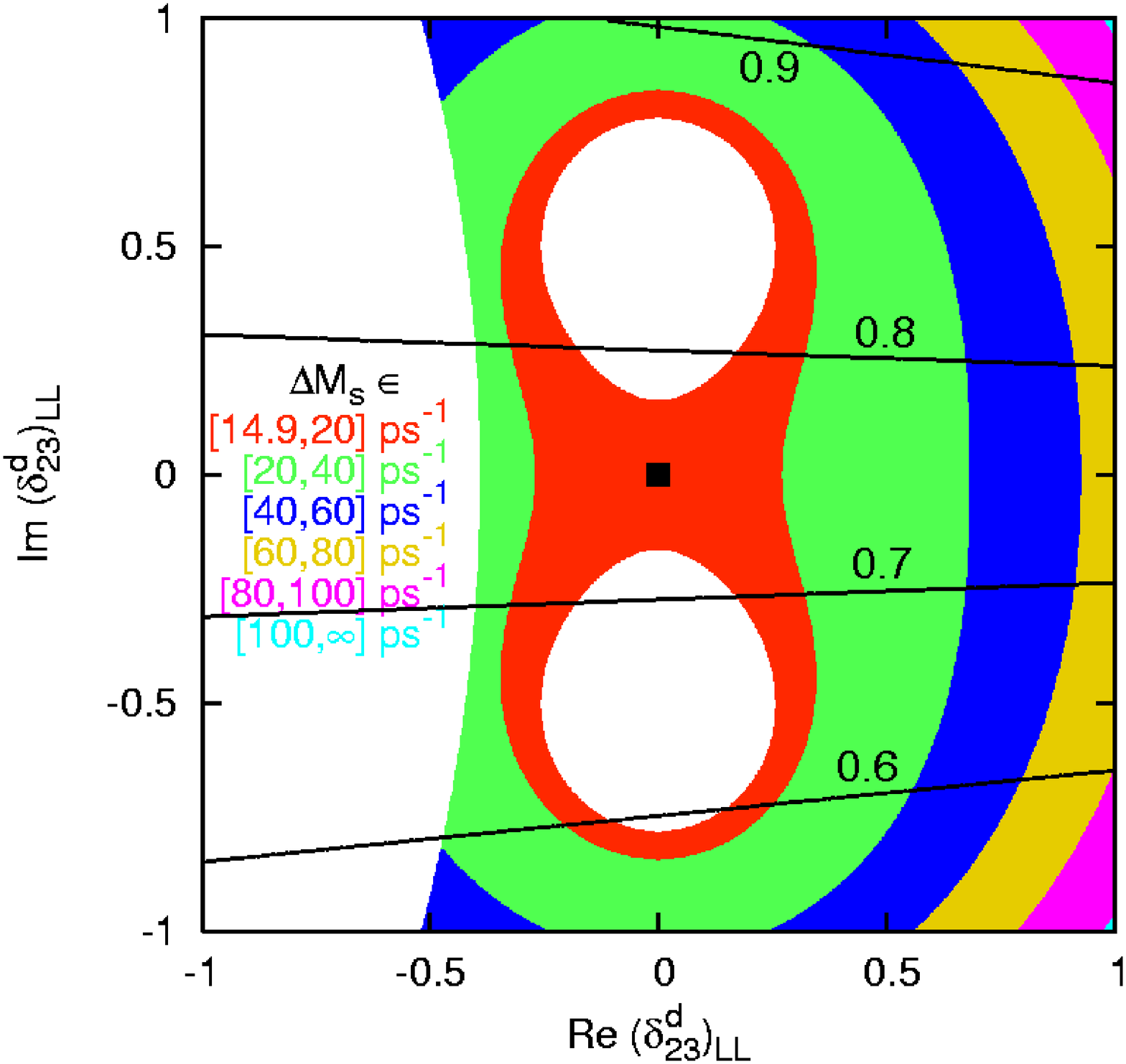}}
{\includegraphics[width=5.0cm]%
{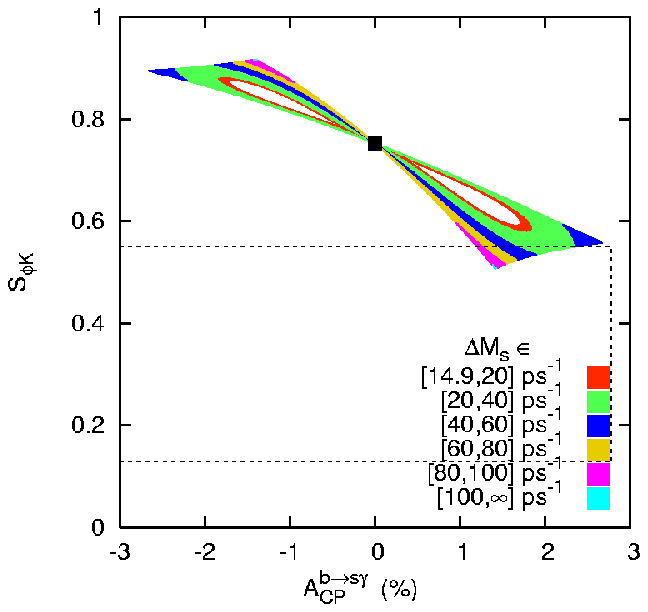}}
{\includegraphics[width=5.0cm]
%{../../susyb/b2phik/400-1_fin/kolda/Ssin2betas-LL.eps}}
{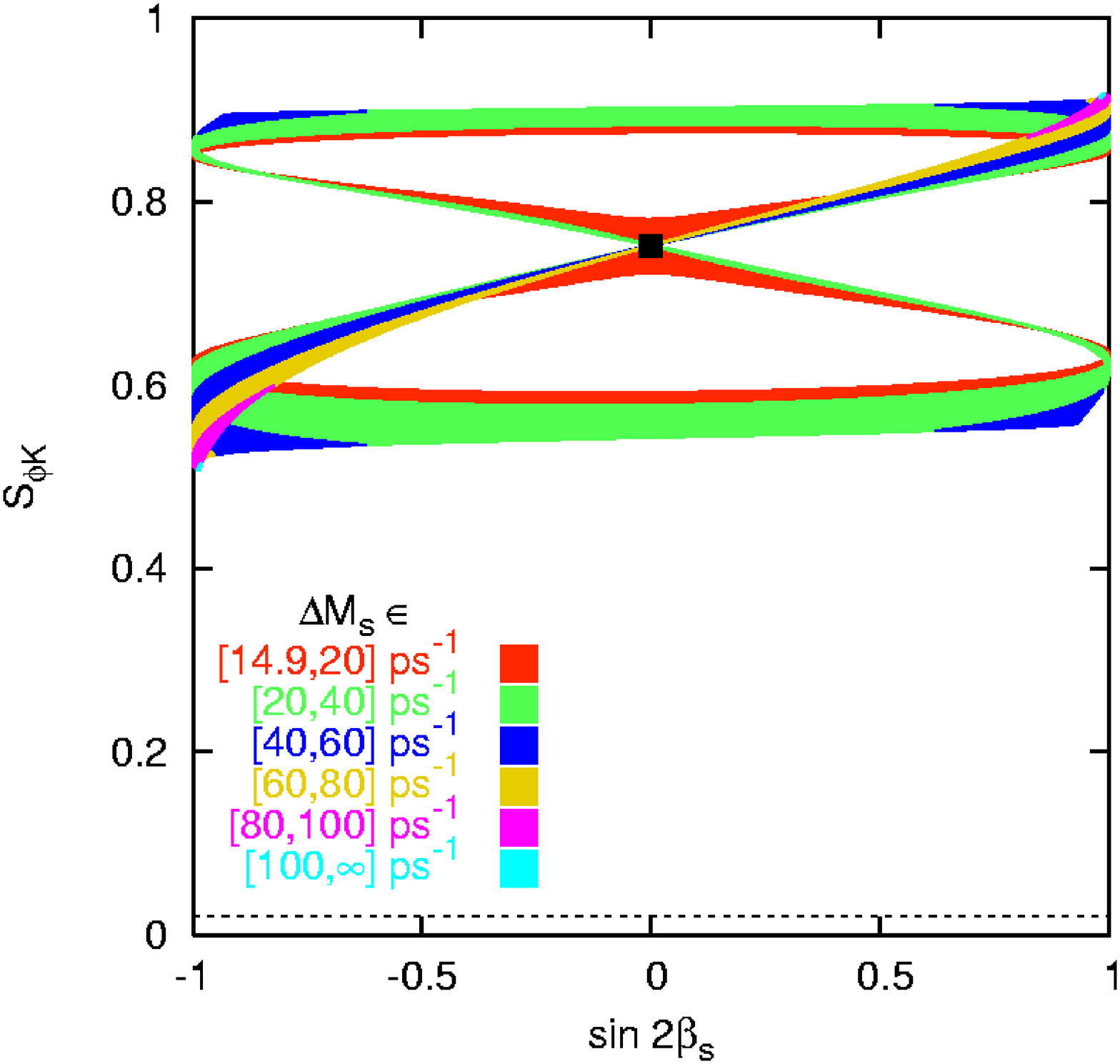}}
 \caption{
The allowed region in the plane of  
(a) the (Re $\delta_{LL}$, Im $\delta_{LL}$ ), 
(b) $S_{\phi K}$ and $A_{\rm CP}^{b\rightarrow s \gamma}$,
(c)  $S_{\phi K}$ and $\sin 2 \beta_s$  
for the case of a single $LL$ insertion,  with
$m_{\tilde{g}} =\tilde m= 400\,$GeV.
The dotted boxes show the current $1\sigma$ expermental bounds, and 
the hahsed regions correspond to 
$B ( B_d \rightarrow \phi K^0 ) > 1.6 \times 10^{-5}$.
 }
\label{fig:LL}
\end{figure}

We begin by considering the case of a single $LL$ mass insertion: 
$(\delta^d_{LL})_{23}$. The results are shown in Fig.~\ref{fig:LL} 
(a)--(c). 
We get similar results for a single $RR$ insertion (see Ref.s~
\cite{kkkpww1,kkkpww2} for more details). 
Scanning over the parameter space consistent with $B\rightarrow X_s \gamma$
and $\Delta M_s$ constraints  (Fig.~\ref{fig:LL} (a)),
we find that $S_{\phi K} > 0.5$
for $m_{\tilde{g}} = \tilde{m} = 400\,$GeV  and 
for any value of $| ( \delta_{LL}^d )_{23} | \leq 1$,
the lowest values being achieved only for very large $\Delta M_s$
(Fig.~\ref{fig:LL} (b) and (c)). 
If we lower the gluino mass down to $250\,$GeV,  $S_{\phi K}$ can shift down 
to $\sim 0.05$, but  only in a small corner of parameter space.
Similar results hold for a single $RR$ insertion.   Thus we 
conclude that the effects of the $LL$ and $RR$ insertions on  
$B\rightarrow X_s \gamma$ and $B\rightarrow \phi K$ are not very dramatic,
although it can marginally accommodate the current world average of 
$S_{\phi K}$. 
Especially it is not likely to generate a negative $S_{\phi K}$,
unless  gluino and squarks are relatively light. Nonetheless, their effects 
on $B_s - \overline{B_s}$ mixing could be very  large, providing a clear 
signature for $LL$ or $RR$ mass insertions (Fig.~\ref{fig:LL} (c)). 

Next we consider the case of a single $LR$ insertion.
Scanning over the parameter space and imposing the constraints 
from $B\to X_s\gamma$ and $\Delta M_s$, we find 
$| ( \delta_{LR}^d )_{23} | \lesssim 10^{-2}$.  This is, however,
large enough to significantly affect $B_d \to\phi K_S$,
both its branching ratio and CP asymmetries, through the
contribution to the chromomagnetic dipole moment operator. %$C_{8g}$.
\begin{figure}
{\includegraphics[width=5.5cm]
%{../../susyb/b2phik/400-1_fin/kolda/deltaLR.eps}}
{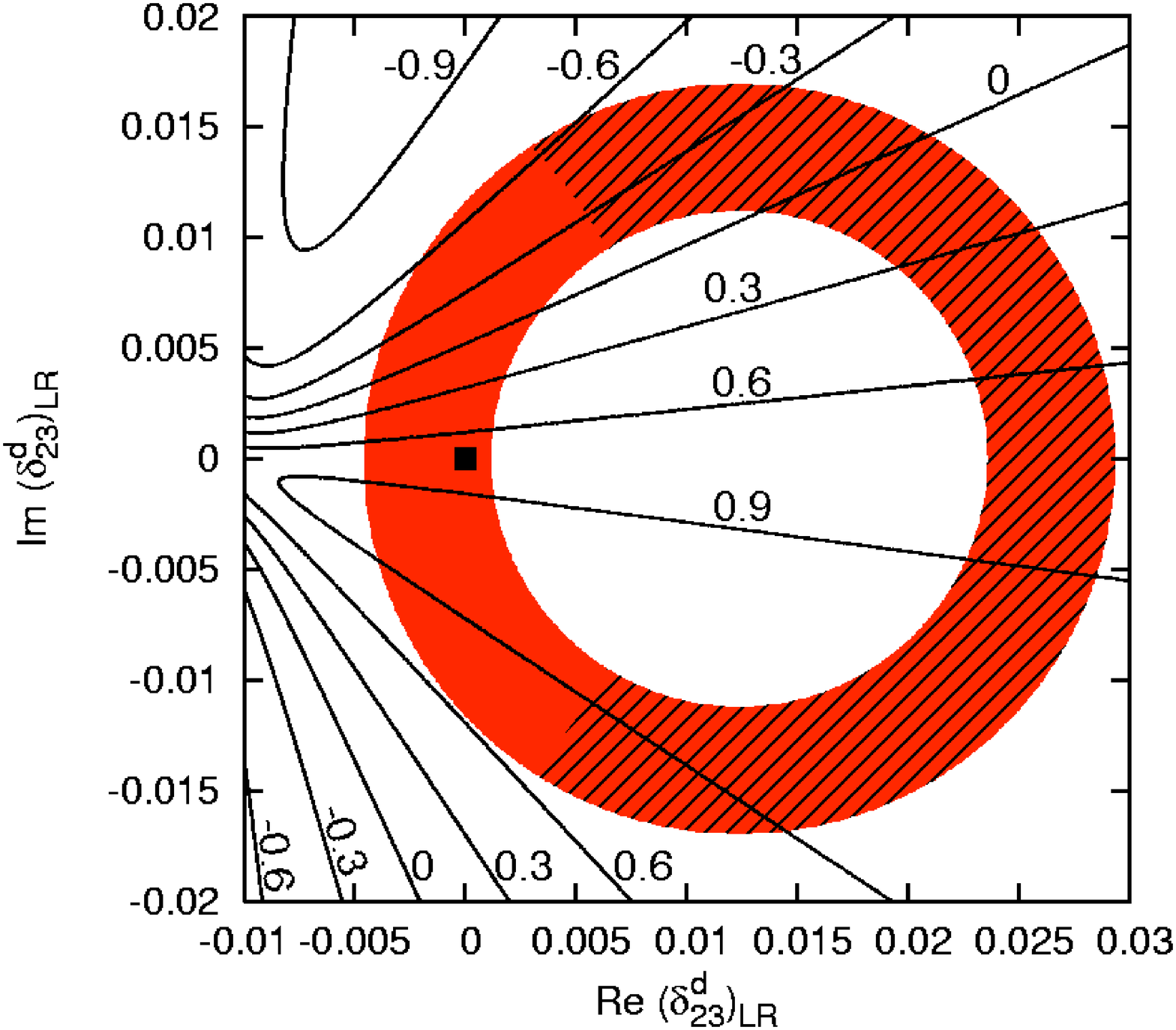}}
{\includegraphics[width=5.0cm]%
{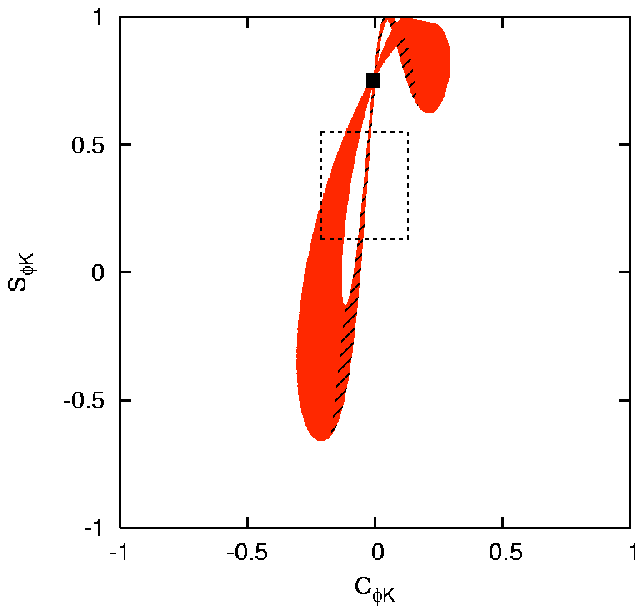}}
{\includegraphics[width=5.0cm]
{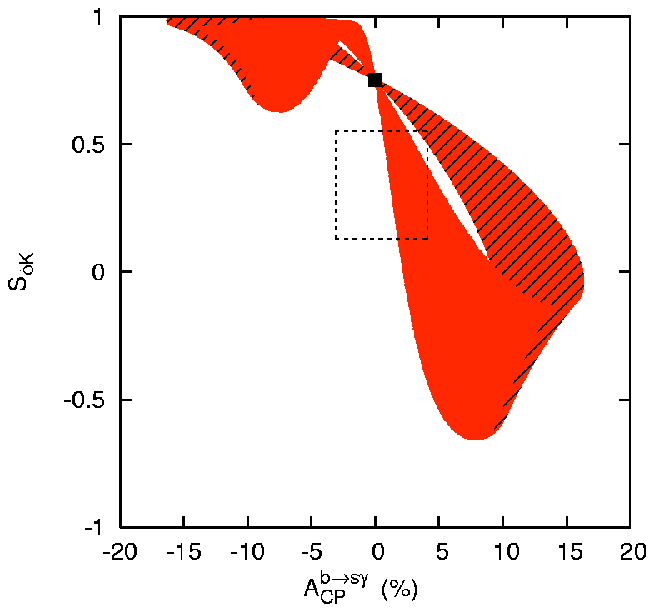}}
 \caption{
The allowed region in the plane of  
(a) the (Re $\delta_{LR}$, Im $\delta_{LR}$ ), 
(b) $S_{\phi K}$ and $C_{\phi K}$, and  
(c) $S_{\phi K}$ and $A_{\rm CP}^{b\rightarrow s \gamma}$,
for the case of a single $LR$ insertion,  with
$m_{\tilde{g}} =\tilde m= 400\,$GeV.
The dotted boxes show the current $1\sigma$ expermental bounds, and 
the hashed regions correspond to 
$B ( B_d \rightarrow \phi K^0 ) > 1.6 \times 10^{-5}$.
 }
\label{fig:LR}
\end{figure}
In Fig.s~\ref{fig:LR} (a) and (b), we show the allowed region in the complex 
$( \delta_{23}^d )_{LR}$ plane with the contours of $S_{\phi K}$, and 
the correlation between $S_{\phi K}$ and $C_{\phi K}$.
Since the  $LR$ insertion can have a large effect on the
CP-averaged branching ratio for $ B\rightarrow \phi K$ 
we further impose that $B ( B\rightarrow \phi K) < 1.6 \times 10^{-5}$ 
(which is twice the experimental value)
in order to include theoretical uncertainties in the BBNS approach 
related to hadronic physics. We can see that the 
$B\rightarrow \phi K$ branching ratio constrains 
$ ( \delta_{LR}^d )_{23} $ just as much as $B\rightarrow X_s \gamma$.
Also we can get a large negative $S_{\phi K}$, but only if $C_{\phi K}$ 
is also negative. 
The correlation between $S_{\phi K}$ and the direct CP asymmetry in 
$B \rightarrow X_s \gamma$ ($\equiv A_{\rm CP}^{b\rightarrow s \gamma}$) 
is shown in Fig.~\ref{fig:LR} (c). 
We find $A_{\rm CP}^{b\rightarrow s \gamma}$ 
becomes positive for a negative $S_{\phi K}$, while a negative 
$A_{\rm CP}^{b\rightarrow s \gamma}$ implies that $S_{\phi K} > 0.6$.  
The present world average  on $A_{\rm CP}^{b\rightarrow s \gamma}$ 
gives additional  constraint on the $LR$ model, and the resulting 
$S_{\phi K}$ is consistent with the data.  Finally, 
the deviation of $B_s - \overline{B}_s$ mixing from the SM prediction 
is very small for $| ( \delta_{23} )_{LR} | \lesssim 10^{-2}$.
Thus we conclude that a single $LR$ insertion can accommodate 
large deviation in $S_{\phi K}$ from the SM rather easily with 
$\tilde{m} =  m_{\tilde{g}} = 400$ GeV. 
This scenario can be tested by measuring a positive direct CP asymmetry 
in $B\to X_s\gamma$ and $B_d$-$\bar B_d$ mixing consistent with the SM. 

We also studied the $RL$ dominance scenario, and the generic feature is
similar to the $LR$ insertion case except that (i) the 
$B\rightarrow X_s \gamma$ branching ratio gives a different constraint  
from the $LR$ insertion case, since the SM contribution does not interfere
with the $RL$ contribution, and (ii) direct CP asymmetry in 
$B\rightarrow X_s \gamma$ is zero unless there is additional $RR$ insertion. 
See Ref.s~\cite{kkkpww1,kkkpww2} for further detail.

Now let us provide possible motivation for values of
$( \delta_{LR,RL}^d )_{23} \lesssim 10^{-2}$ that could shift $S_{\phi K}$ 
from the SM value rather easily.  In particular, 
at large $\tan\beta$ it is possible to have double mass insertions
which give sizable contributions to $( \delta_{LR,RL}^d )$. First a
$( \delta_{LL}^d )_{23}$ or $( \delta_{RR}^d )_{23}\sim 10^{-2}$ is
generated. The former can be obtained from renormalization group
running even if its initial value is negligible at the high scale.
The latter may be implicit in SUSY GUT models with large mixing in the
neutrino sector \cite{moroi}.
Alternatively, in models in which the SUSY flavor problem is resolved 
by an alignment mechanism using 
spontaneously broken flavor symmetries, or  by decoupling,  
the resulting $LL$ or $RR$ mixings in the $23$ sector could easily be of 
order $\lambda^2$ \cite{align,decoupling}.  However as discussed above, 
this size of the $LL$ and/or $RR$ insertions can not explain
the measured CP asymmetry in $B_d \rightarrow \phi K_S$ unless the squarks 
and gluinos are rather light.
But  at large $\tan\beta$, the $LL$ and $RR$ insertions can
induce the $RL$ and $LR$ insertions needed for $S_{\phi K}$ through a  
double mass insertion \cite{ko1,ko2}:
\[
( \delta_{LR,RL}^d )_{23}^{\rm ind} =  ( \delta_{LL,RR}^d )_{23} \times 
{ m_b ( A_b - \mu \tan\beta ) \over \tilde{m}^2 }.
\]
One can achieve $( \delta_{LR,RL}^d )_{23}^{\rm ind} \sim 10^{-2}$
if $\mu \tan\beta \sim 10^4\,$GeV, which could be natural if $\tan\beta$ is 
large (for which $A_b$ becomes irrelevant). 
Note that in this scenario both the $LL (RR)$ and $LR (RL)$ insertions 
would have the same CP violating phase, since the phase of $\mu$ here is
constrained by electron and down-quark electric dipole moments. 
Lastly, one can also construct string-motivated $D$-brane 
scenarios in which $LR$ or $RL$ insertions are $\sim 10^{-2}$ \cite{kkkpww2}. 

Summarizing this section, we considered several classes of  potentially 
important SUSY contributions to $B\rightarrow \phi K_S$  in order to 
see if a significant deviation in its time-dependent CP asymmetry 
$S_{\phi K}$ could arise from SUSY effects. The Higgs-mediated FCNC effects
are small. The models based on the gluino-mediated $LL$ and $RR$ insertions 
give a rather small deviation in $S_{\phi K}$ from the SM prediction,
unless the squarks and gluinos are relatively light. 
On the other hand, the gluino-mediated contribution with $LR$ and/or $RL$ 
insertions can lead to sizable deviation in $S_{\phi K_S}$, as long as 
$| ( \delta_{LR,RL}^d )_{23} | \sim 10^{-3} - 10^{-2}$.
As a byproduct, we found that nonleptonic $B$ decays such as $B\rightarrow
\phi K$ begin to constrain $| ( \delta_{LR,RL}^d )_{23} | $ 
as strongly as $B\rightarrow X_s \gamma$.  
Besides producing no measurable deviation in $B^0 - \bar B^0$ mixing, 
the $RL$ and $LR$  operators generate definite
correlations among $S_{\phi K}$, $C_{\phi K}$ and 
$A_{\rm CP}^{b\rightarrow s \gamma}$, and our prediction for 
$S_{\phi K}$ can be easily tested by measuring these other  observables. 
Finally, we also point out that the $| ( \delta_{LR,RL}^d )_{23} | 
\lesssim 10^{-2}$ can be naturally obtained in SUSY flavor models with 
double mass insertion at large $\tan\beta$, and in 
string-motivated models \cite{kkkpww2}.

%%%%%%%%%%%%%%%%%%%%%%%%%%%%%%%%%%%%%%%%%%%%%%%%%%%%%%%%%%%%%%%%%%%%%%%%%%
\section{$s_R \rightarrow d_L$ transition induced by 
$b_R \rightarrow s_R$ transition
: Re ($\epsilon^{'}/\epsilon_K$) vs.  $S_{\phi K}$}

Assuming that the current low value of $S_{\phi K}$ is a signal of new 
physics, we need a new CP violating phase in $b\rightarrow s$ transition. 
An attractive possibility for such new physics beyond the SM occurs in  
supersymmetric grand unified theories (SUSY GUT's)  scenarios with 
seesaw mechanism for neutrino masses and mixings. In such scenarios, 
the large atmospheric neutrino oscillation can be related with a
large $b\rightarrow s$ transition through down type squark and gluino loop 
effects. This flavor changing effect is parametrized by 
a mixing parameter $( \delta^d_{23} )_{RR}$ with a CP phase $\sim O(1)$. 
For low $\tan\beta$, the single $RR$ insertion can lead to some deviation
in $S_{\phi K}$, if gluinos and squarks are relatively light.
For large $\tan\beta$ case, the double mass insertion can lead to effective
$RL$ insertion of $10^{-2}$, leading to a significant deviation 
in $S_{\phi K}$ from the SM prediction.
In Fig.~5 (a), we show the Feynman diagram for $b\rightarrow s g$ 
involving a CP violating $( \delta^d_{23} )_{RR}$. 

The new CP violating phase in the $RR$ insertion can affect the strange 
quark chromo-electric dipole moment (CEDM) through triple mass insertions, 
if there is an $LL$ insertion between 
third and second generation down squarks.  
[ Fig.~5 (b) ] \cite{Hisano:2004tf,Kane:2004ku,Abel:2004te}. 
Since the $LL$ insertion is generically present in the minimal 
supergravity (mSUGRA) case,  the strange quark CEDM puts a strong 
constraint on the possible deviation of $S_{\phi K}$ from the SM prediction. 
However, the substantial theoretical uncertainties occurring when one  
relates the quark CEDM's with the hadronic EDM's suggest that 
it would be preferable to have some other observable at disposal 
in addition to the strange quark CEDM in order to constrain $S_{\phi K}$.

In this section, we point out that the  phase in the 
$( \delta_{23}^d )_{RR}$ mixing parameter that would affect $S_{\phi K}$ 
can also contribute to direct CP violation within the neutral kaon system, 
namely Re ($\epsilon^{'} / \epsilon_K$) through triple mass insertion.
The  Feynman diagram for Re ($\epsilon^{'} / \epsilon_K$) with triple 
mass insertion [ Fig.~5 (c) ] is very similar to the Feynman diagram for 
the strange quark CEDM [ Fig.~5 (b) ].  
Needless to say, making use of the observable Re ($\epsilon^{'} / 
\epsilon_K$) to constrain some SUSY soft breaking parameters also entails 
theoretical uncertainties mainly ascribed to our ignorance in the 
evaluation of the relevant hadronic matrix elements. Our discussion shows 
that, even taking into account such a huge degree of uncertainty, 
Re ($\epsilon^{'} /\epsilon_K$) still constitutes a precious tool in 
constraining the interesting flavor changing  mass insertion parameter 
$( \delta^d_{32} )_{RR}$ and, in 
any case, it plays at least a complementary role to the strange quark CEDM 
in performing such a task.   

\begin{figure}
\includegraphics[width=5cm]{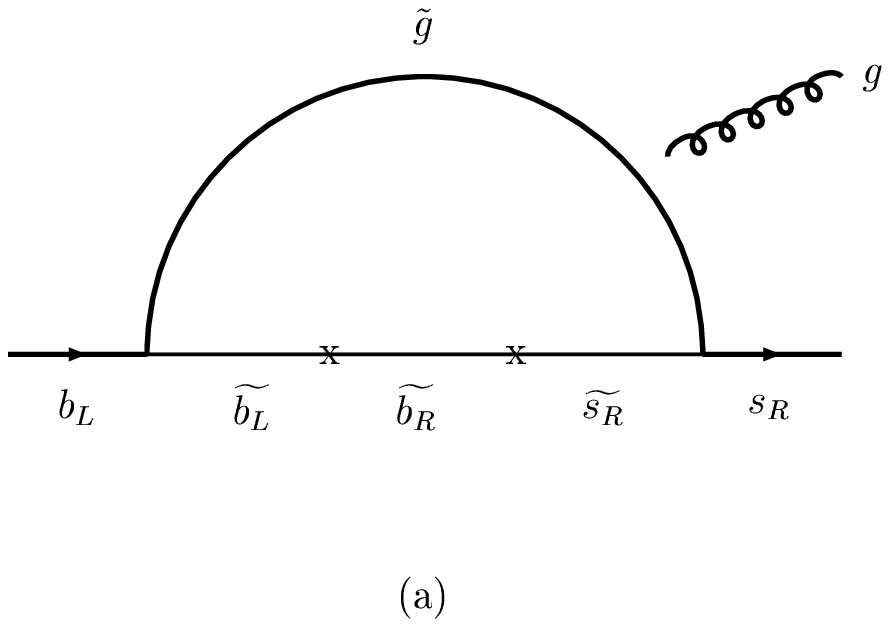} %\\
%\vspace{1cm}
\includegraphics[width=5cm]{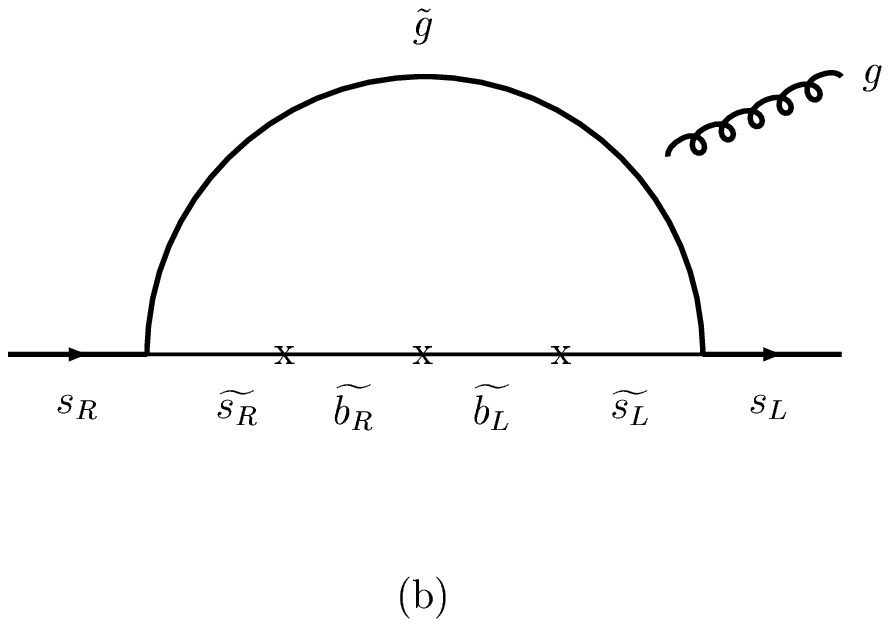} %\\
%\vspace{1cm}
\includegraphics[width=5cm]{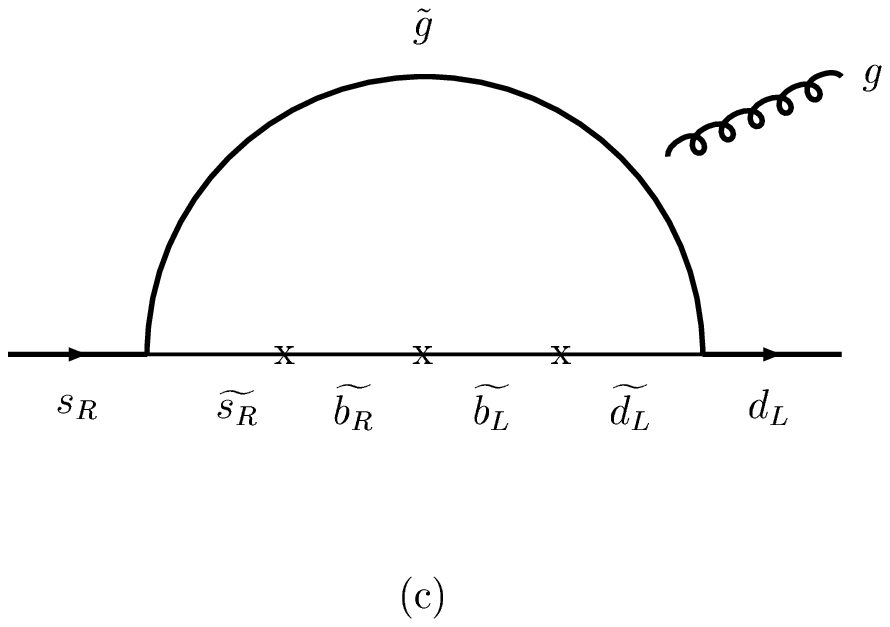}%
\caption{\label{} Feynman diagrams for  
(a) $b\rightarrow s g$ with double mass insertions,
(b) strange quark CEDM with triple mass insertions,
and (c) $s\rightarrow d g$ with triple mass insertions,
%and (c) strange quark CEDM with triple mass insertions, 
involving  $( \delta^d_{23} )_{RR}$ as the dominant source of
new  CP violating parameter contributing to $S_{\phi K}$ and
{\rm Re} ($\epsilon^{'} / \epsilon_K$).
}
\end{figure}

In this talk, I'll illustrate the main point in a simple way, relegating 
the detailed  analysis including theoretical uncertaintiess to 
Ref.~\cite{kmp}. 
If the $( \delta^d_{23} )_{RR}$ mixing is the dominant new physics 
contribution to $B_d \rightarrow \phi K_S$, we find the following from 
(5) and (9) in the previous sections :
\begin{eqnarray}
{\rm Re} (\epsilon^{'}/\epsilon_K) : 
C_{8g}^{\rm SUSY} ( \Delta S = 1 ) & \propto & 
f_{1} (x)~\de{13}{LL} \de{33}{LR} \de{32}{RR} ,   
\\
S_{\phi K} : C_{8g}^{\rm SUSY} ( \Delta B = 1 ) & \propto & 
f_{2} (x)~\de{32}{RR} + f_{3} (x)~ { m_{\tilde{g}} \over m_b}~
\de{33}{RL} \de{23}{RR} ,
\end{eqnarray}
where $f_{i=1,2,3} (x)$ are the loop functions obtained in the previous
sections. 
Now, if the SUSY contribution saturates Re ($\epsilon^{'}/\epsilon_K$), 
then  it is well known that  one has to satisfy 
\[
| \de{13}{LL} \de{33}{LR} \de{32}{RR} | \lesssim 10^{-5}
\]
with an $O(1)$ phase \cite{Gabbiani:1996hi}.
Since the RG evolution generates $\de{13}{LL} \sim \lambda^3$ within 
mSUGRA scenario, we can  derive the following upper bound:
\begin{equation}
\label{eq:constraint}
| \de{33}{LR} \de{32}{RR} | \lesssim 10^{-2}.   
\end{equation}
Note that this combination enters the calculation
of $S_{\phi K}$ and $B\rightarrow X_s \gamma$ 
through $C_{8g(7\gamma)} ( \Delta B = 1)$ along with $\de{32}{RR}$.
For a small $\mu\tan\beta$ (corresponding to a small $\de{33}{RL}$), 
one can have larger $\de{32}{RR}$, which is constrained by the lower bound 
on $\Delta M_s$  and the $B\rightarrow X_s \gamma$ branching ratio. 
For a large $\mu \tan\beta$ (corresponding to a large $\de{33}{RL}$), 
$\de{32}{RR}$ should be smaller in order to satisfy (\ref{eq:constraint}), 
which would be less constrained by
$\Delta M_s$  and the $B ( B\rightarrow X_s \gamma)$. 
In either case, we can expect that the deviation in $S_{\phi K}$ cannot 
be that large for such $\de{32}{RR}$  satisfying the 
Re ($\epsilon^{'}/\epsilon_K$) constraint, (\ref{eq:constraint}). 

In Figs.~6 (a) and (b), we show  the  plots for $S_{\phi K}$ and 
Re ($\epsilon^{'} / \epsilon_K$) for $\mu \tan \beta = 1$ and $5$ TeV, 
respectively,  with $\tilde{m} = m_{\tilde{g}} = 500$ GeV.
The thick vertical error bar shows the current data on  $S_{\phi K}$, 
and the two dashed vertical lines delimit the experimental value
of Re ($\epsilon^{'} / \epsilon_K$) \cite{Eidelman:2004wy},
\begin{equation}
  \mathrm{Re}\ (\epsilon^{'} / \epsilon_K)
  = ( 16.7 \pm 2.6 ) \times 10^{-4} .
\end{equation}
The full black box shows our estimation of $S_{\phi K}$ and
Re ($\epsilon^{'} / \epsilon_K$) within the SM.  
Its width and height are the uncertainties
in Re ($\epsilon^{'} / \epsilon_K$) and $S_{\phi K}$, respectively.
Note that the Re ($\epsilon^{'} / \epsilon_K$) data
gives a strong constraint on the possible value of $S_{\phi K}$ 
even if there are large hadronic uncertainties.
\begin{figure}
\begin{center}
%\subfigure[]%
%{\includegraphics[width=8cm]{Sepspeps-RRRL-uncer+-1-1.eps}}%
\subfigure[$\mu \tan \beta = 1$ TeV]%
{\includegraphics[width=6cm]{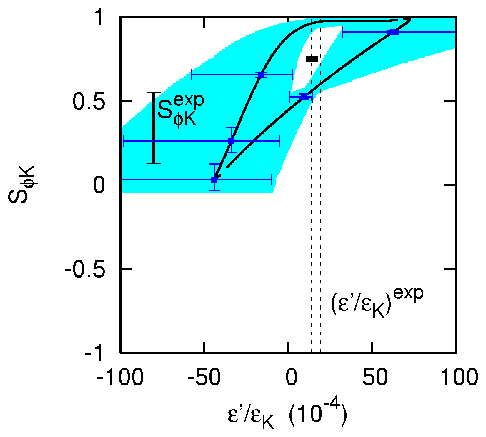}}
%\subfigure[]%
%{\includegraphics[width=8cm]{Sepspeps-RRRL-uncer+-5-1.eps}}%
\subfigure[$\mu \tan \beta = 5$ TeV]%
{\includegraphics[width=6cm]{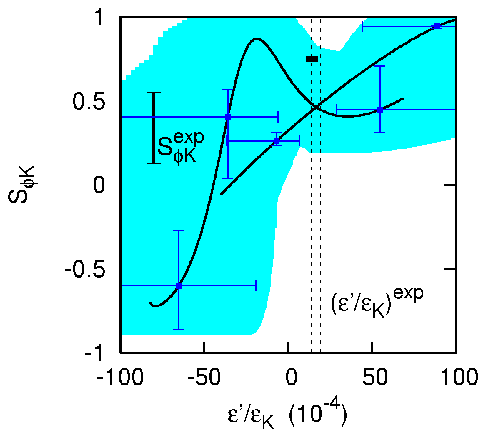}}%
\end{center}
\caption{
%\label{fig1} 
$S_{\phi K}$ vs. {\rm Re} ($\epsilon^{'} / \epsilon_K$) for
(a) $\mu \tan \beta = 1$ TeV and (b) $\mu \tan \beta = 5$ TeV.
with $\tilde{m} = m_{\tilde{g}} = 500$ GeV.
Experimental bounds on {\rm Re} ($\epsilon^{'} / \epsilon_K$)
and $S_{\phi K}$ are depicted by the vertical dashed lines
and the thick vertical error bar, respectively.
The SM predictions of them
are marked by the black box, whose extent indicates
their uncertainties.
The black curve does not include hadronic uncertainties, and
the gray region includes them.
The respective uncertainties in {\rm Re} ($\epsilon^{'} / \epsilon$)
and $S_{\phi K}$ are shown by the horizontal and
vertical error bars at some selected points.
}
\end{figure}
Note that the constraint from Re ($\epsilon^{'} / \epsilon_K$) is
comparable to that from the strange quark CEDM.
In particular the positive (negative)  $S_{\phi K}$ is correlated 
with the positive (negative) Re ($\epsilon^{'} / \epsilon_K$) 
within minimal SUGRA boundary conditions. In particular, the old Belle
data with the negative $S_{\phi K}$ implies a negative
Re ($\epsilon^{'} / \epsilon_K$) in the $RR$ dominance scenario such as 
SUSY GUT models with right-handed neutrinos, which is clearly excluded 
by the data Re $(\epsilon^{'}/\epsilon_K) = (16.7 \pm 2.6 ) \times 10^{-4}$.  
If the old Belle data were still valid, then the $RR$ dominace scenario
should have been discarded. 
Our results provide a meaningful correlation between $S_{\phi K}$ and
Re ($\epsilon^{'} / \epsilon_K$) despite of large hadronic uncertainties
in both quantities. This is independent of the strange  quark CEDM 
constraint, and probably has less theoretical uncertainties.

If we considered more general flavor structures in the  scalar masses 
at $M_*$,  then our results will be changed accordingly. 
The Wilson coefficients for $C_{8g}$'s for both $\Delta B (S) =1$ have
to include other mass insertion parameters such as $\de{23}{LR}$,
$\de{12}{LL}$, $\de{23}{LL}$, etc., which were neglected in Secs.~II and 
III because they are small within the mSUGRA scenarios. Still we should 
make it sure that  the new flavor physics that affects $S_{\phi K}$ 
does not contribute to Re($\epsilon^{'}/\epsilon_K$) too much, and 
this could make a strong constraint on new sources of flavor and CP 
violation despite of theoretical uncertainties on 
Re($\epsilon^{'}/\epsilon_K$). 

In summary, we showed that if the $RR$ $b\rightarrow s$ transition is
large with $O(1)$ phase, it can affect not only $S_{\phi K}$ through
the double mass insertion and the strange quark CEDM through triple
mass insertion, it affects also the Re ($\epsilon^{'} / \epsilon_K$).
The correlation between the two observables are  strong despite large
hadronic uncertainties in both observables within mSUGRA boundary 
conditions with flavor universal scalar masses at $M_*$. 
The current data on Re ($\epsilon^{'} / \epsilon_K$) indicates that
$S_{\phi K}$ should be in the range of $0.25$--$1.0$, which is now 
in accord with the present world average of $S_{\phi K}$.

%%%%%%%%%%%%%%%%%%%%%%%%%%%%%%%%%%%%%%%%%%%%%%%%%%%%%%%%%%%%%%%%%%%%%%%%%
\section{$B_s \rightarrow \mu^+ \mu^-$ and SUSY breaking mechanisms}

The Higgs sector of the MSSM is not Type II but Type III two-Higgs doublet
model due to the presence of the soft SUSY breaking terms.  Therefore 
there are loop induced nonholomorphic  trilinear couplings, and this term 
can induce new FCNC involving neutral Higgs bosons \cite{kolda}.   
In the large $\tan\beta$ region, this effect on the $b-s-$Higgs couplings 
can be enhanced by $\tan^2 \beta$, and could dominate the 
$B_s \rightarrow \mu^+ \mu^-$ process within SUSY models in the large
$\tan\beta$ region. Since its branching ratio within the SM is very small 
($(3.7 \pm 1.2) \times 10^{-9}$), this decay mode could be
a sensitive probe of SUSY models in the large $\tan\beta$ region. 
In Refs.~\cite{baek_ko_song}, we studied the correlations between 
$B_s \rightarrow \mu^+ \mu^-$ branching ratio, the muon $(g-2)$, 
and other observables in the $B$ system, imposing the direct search limits 
on Higgs an SUSY particle masses, and $B\rightarrow X_s \gamma$ branching 
ratio and assuming that $(g-2)_\mu^{\rm SUSY} > 0$ (namely $\mu > 0$). 
In this section, I report the main results of Refs.~\cite{baek_ko_song}. 
(The correlation between $(g-2)_\mu$ and $B_s \rightarrow \mu^+ \mu^-$ was
first noticed in Ref.~\cite{dreiner} within the minimal supergravity 
scenario.)

The soft SUSY breaking parameters at electroweak scale is determined by
RG evolution with the initial condition at the messenger scale 
$M_{\rm mess}$ within a given SUSY breaking scenario. 
The initial conditions depend on SUSY breaking mediation
mechanisms: supergravity (including scenarios motivated by superstring 
theories, $M-$theories and $D-$brane models), gauge mediation (GMSB), 
anomaly mediation (AMSB), gaugino mediation, to name a few. Many of these 
scenarios predict flavor blind soft terms at the messenger scale, and 
nontrivial flavor dependence in the soft terms are generated by RG evolution
from $M_{\rm mess}$ to electroweak scale $\mu_{\rm EW}$. Then the dominant 
contribution to $b\rightarrow s$ transition comes from the chargino-stop 
loop diagram. Therefore, in order to have a large branching ratio for 
$B_s \rightarrow \mu^+ \mu^-$, we need large $\tilde{t}_L - \tilde{t}_R$ 
mixing, light chargino and stops, and large $\mu\tan\beta$. 
If these conditions cannot be met, there would be no chance to observe 
$B_s \rightarrow \mu^+ \mu^-$ in the near future at the Tevatron.  

\begin{figure*}[htbp]
\centering
\subfigure[]
{\includegraphics[width=5.5cm,height=5.5cm]
{%../susyb/amsb/prd2/figs2/gmsb-1-15/
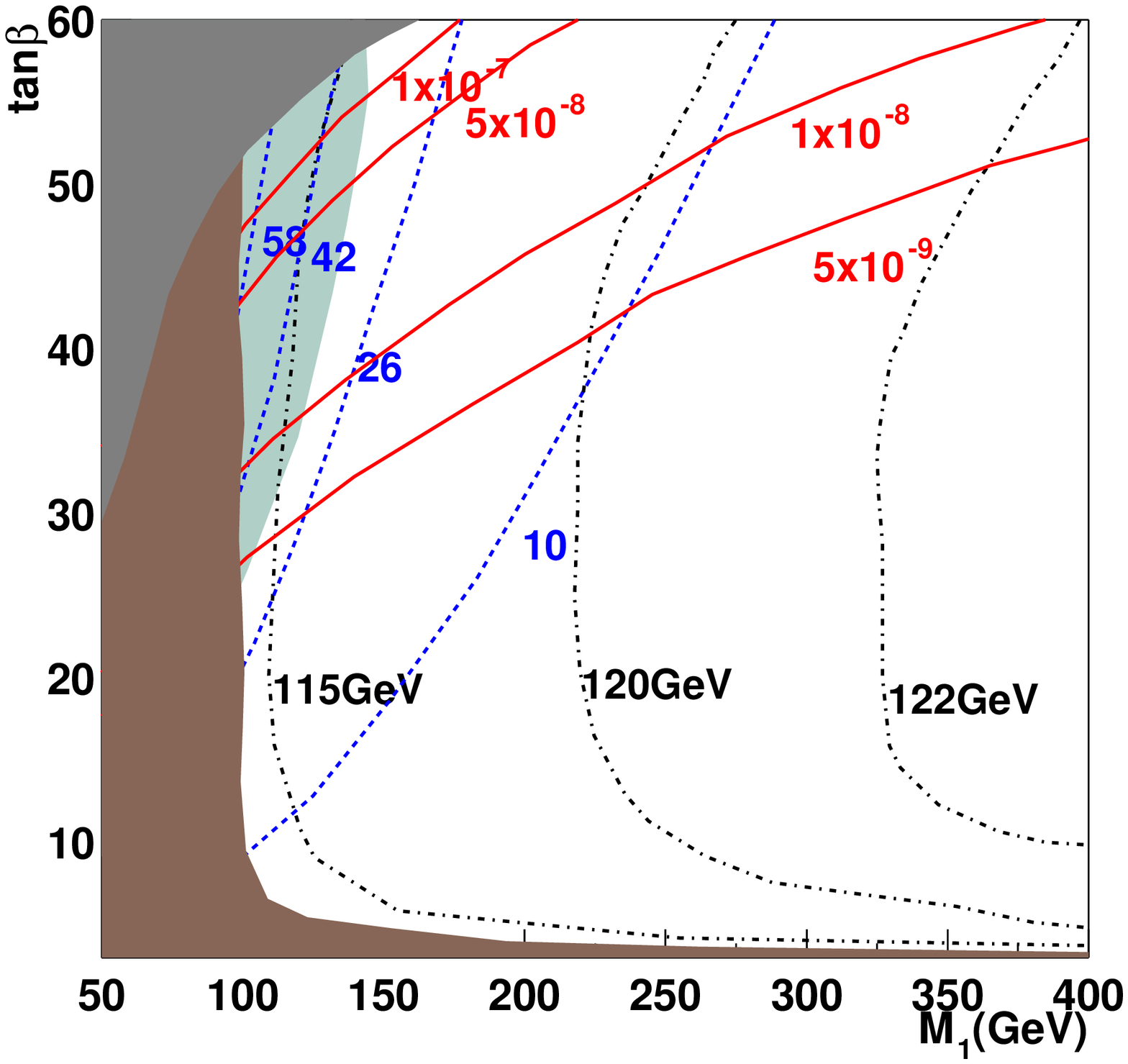}}\hspace*{-5mm}%
\subfigure[]
{\includegraphics[width=5.5cm,height=5.5cm]
{%../susyb/amsb/prd2/figs2/gmsb-5-15/
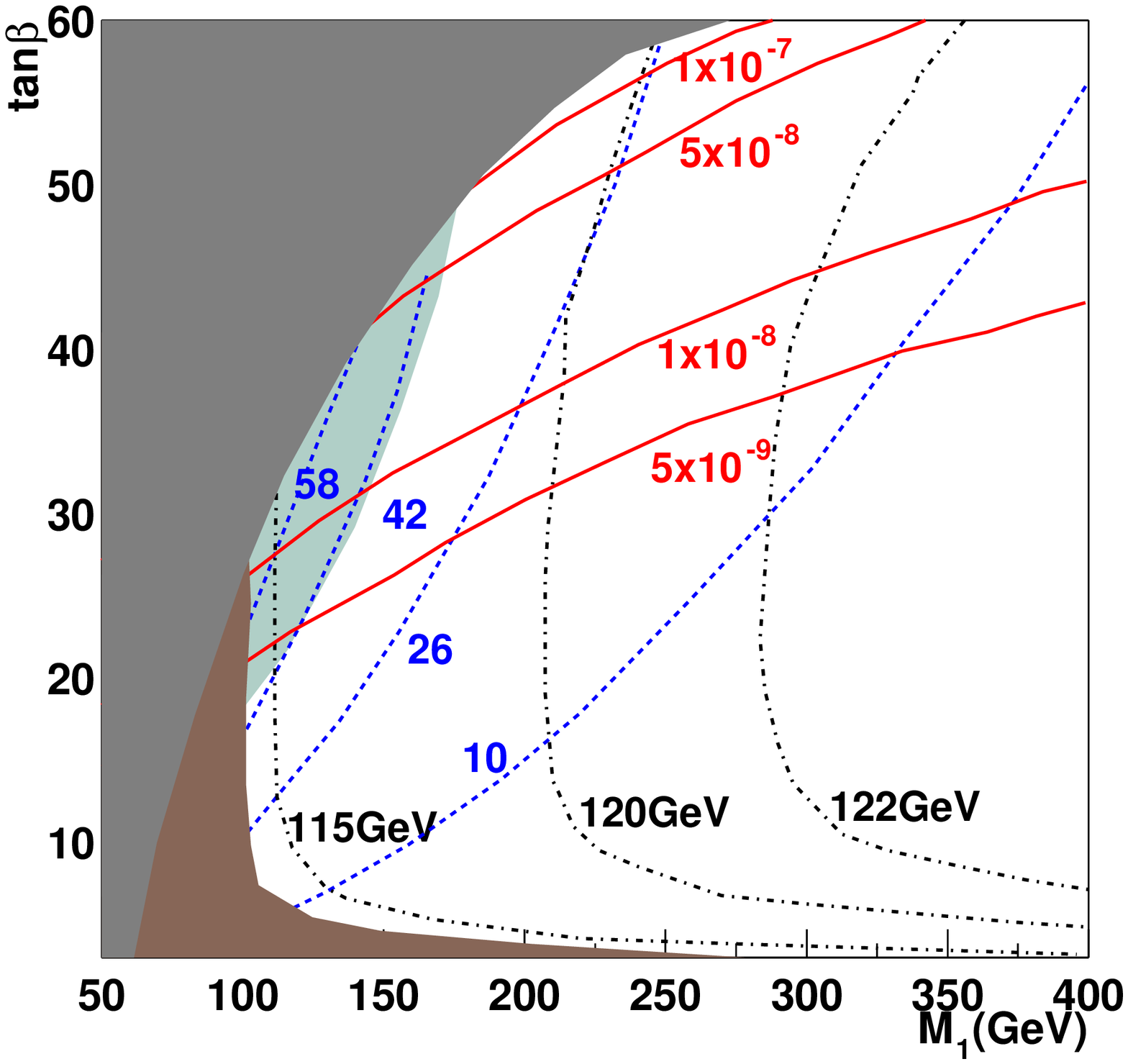}}\hspace*{-2mm}%
\subfigure[]
{\raisebox{1.4mm}{\includegraphics[width=4.8cm,height=4.8cm]
{%../susyb/amsb/amsb-prl2/
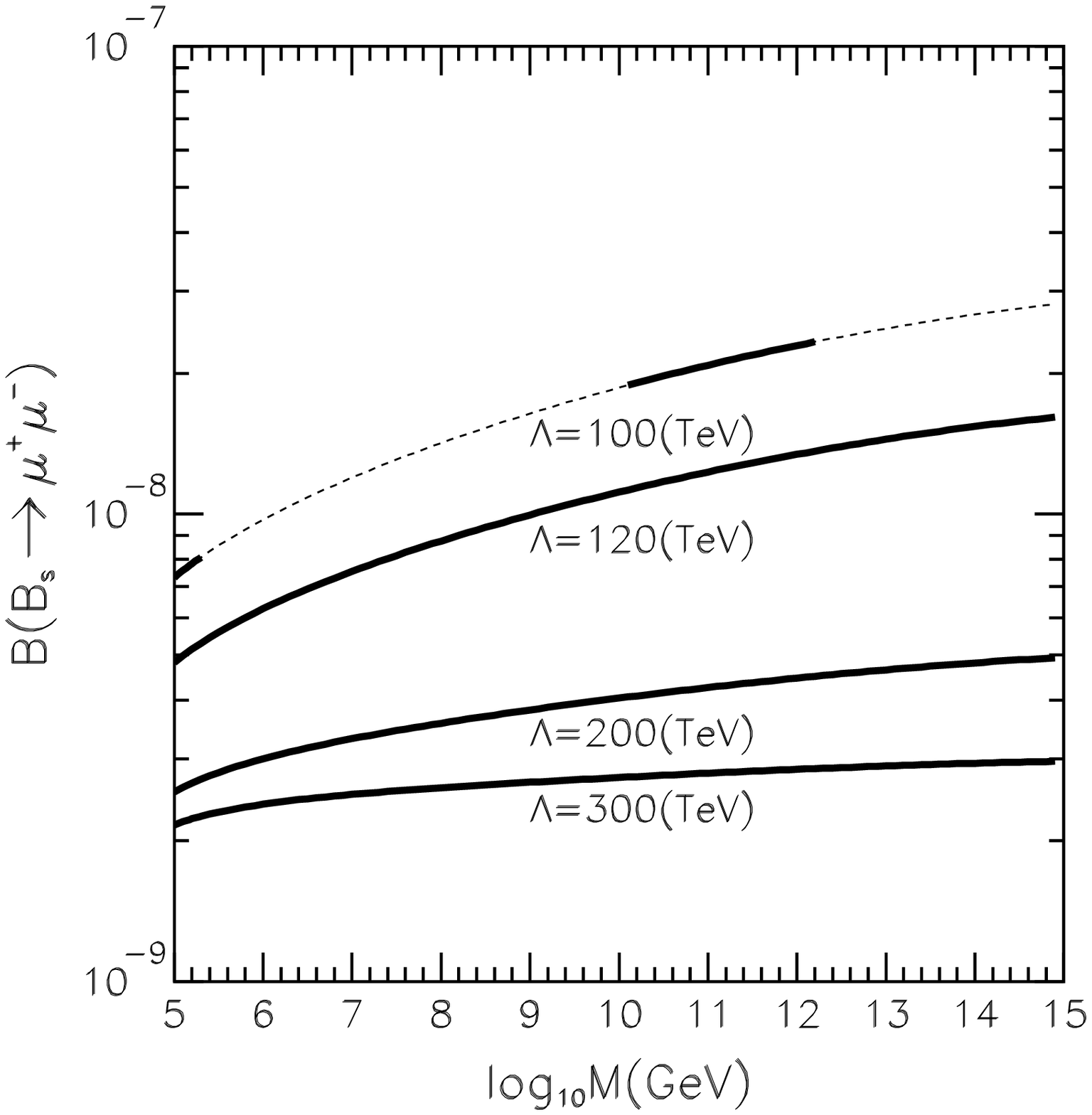}}}
\caption{The contour plots for $a_\mu^{\rm SUSY}$ in unit of $10^{-10}$ 
(in the blue short dashed curves), the lightest neutral Higgs mass (in the 
black dash-dotted curves) and the Br ($B_s \rightarrow \mu^+ \mu^- $) 
(in the red solid curves) for the GMSB scenario in the 
$( M_1, \tan\beta)$ plane with 
(a)  $N_{\rm mess} = 1$ and $M_{\rm mess} = 10^{15}$ GeV, 
(b)  $N_{\rm mess} = 5$ and $M_{\rm mess} = 10^{15}$ GeV. 
In (c), we show the branching ratio for 
$B_s \rightarrow \mu^+ \mu^-$ as a function of 
the messenger scale $M_{\rm mess}$ in the GMSB with $N_{\rm mess} = 1$ for 
various $\Lambda$'s with a fixed  $\tan\beta = 50$. 
The dashed parts are excluded by the direct search limits
on the Higgs and SUSY particle masses. 
}
\label{fig:gmsb}
\end{figure*}

As an example, let us consider GMSB scenarios, which are specified by 
the following set of parameters: 
$M$, $N$, $\Lambda$, $\tan\beta$ and sign($\mu$), where $N$ is the number of 
messenger superfields, $M$ is the messenger scale, and the $\Lambda$ is SUSY 
breaking scale, $\Lambda \approx \langle F_X \rangle / \langle X \rangle$,
where $X$ is a gauge singlet superfield $X$,  the vacuum expectation value 
of which (both in the scalar and the $F$ components) will induce 
SUSY breaking in the messenger sector.
If the messenger scale (where the initial conditions for the 
renormalization group (RG) running for soft parameters are given) is low
such as $10^6 $ GeV, the flavor changing amplitude involving the 
gluino-squark is negligible and only the chargino-upsquark contribution is
important in $B\rightarrow X_s \gamma$.  Also, in the GMSB scenario with low 
messenger scale, the charged Higgs and stops are heavy and their effects on 
the $B\rightarrow X_s \gamma$ and $B_s \rightarrow \mu^+ \mu^-$ are small. 
Also  $A_t$ is small, since it can generated by only RG running. 
Therefore  the stop mixing angle becomes also small. 
These effects lead to very small branching ratio for 
$B_s \rightarrow \mu^+ \mu^-$  ($\lesssim 10^{-8}$), 
making this decay unobservable at the Tevatron  Run II.  
On the other hand, the $a_\mu^{\rm SUSY}$ can be as large as 
$60 \times 10^{-10}$. If we assume the messenger scale be as high as the 
GUT scale, the RG effects become strong and the stops get lighter. 
Also the $A_t$ parameter becomes larger at the electroweak scale, and so is 
the stop mixing angle. 
Therefore the chargino-stop loop contribution can overcompensate the SM and 
charged Higgs - top contributions to $B\rightarrow X_s \gamma$ and this  
constraint becomes more important compared to the lower messenger scale.
Also the $B_s \rightarrow \mu^+ \mu^-$ branching ratio can be enhanced 
(upto $2 \times 10^{-8}$  for $\tan\beta = 50$, for example), 
because stops become lighter and larger $\tilde{t}_L - \tilde{t}_R$ mixing
is possible [ Fig.~\ref{fig:gmsb} (a) ].  
If the number of messenger field is increased from 
$N = 1$ to 5, for example, the scalar fermion masses become smaller 
at the messenger scale,
and stops get lighter in general. Therefore the chargino-stop effects in 
$B\rightarrow X_s \gamma$ and $B_s \rightarrow \mu^+ \mu^-$ get more 
important than the $N=1$ case, and the $B_s \rightarrow \mu^+ \mu^-$ 
branching ratio can be enhanced upto $2 \times 10^{-7}$ 
[ Fig.~\ref{fig:gmsb} (b) ]. 
In short, the overall features in the GMSB scenarios with high messenger 
scale look alike the mSUGRA with $A_0 = 0$. Especially the
branching ratio for the decay $B_s \rightarrow \mu^+ \mu^-$ can be much 
more enhanced for large $\tan\beta$ in the GMSB scenario with high 
messenger scale [ Fig.~\ref{fig:gmsb} (c) ]. 
Thus, if $a_\mu^{\rm SUSY} > 0$ and the decay 
$B_s \rightarrow \mu^+ \mu^-$ is observed at the Tevatron Run II with 
the branching ratio larger than $2 \times 10^{-8}$, the GMSB 
scenario with $N=1$ would be excluded upto $M_{\rm mess} \sim 10^{10}$ GeV
and $\tan\beta \lesssim 50$. 

In the AMSB scenario, the hidden sector SUSY breaking is assumed to be 
mediated to our world only through the auxiliary component of the 
supergravity multiplet (namely super-conformal anomaly) \cite{amsb}. 
In this scenario, the gaugino masses are proportional to
the one-loop beta function coefficient for the MSSM gauge groups, whereas the
trilinear couplings and scalar masses are related with the anomalous 
dimensions and their derivatives with respect to the renormalization scale.
Since the original AMSB model suffers from the tachyonic slepton problem, 
we simply add a universal scalar mass $m_0^2$ to the scalar fermion 
mass parameters of the original AMSB model, and assume that the aforementioned
soft parameters make initial conditions at the GUT scale for the RG evolution.
Thus, the minimal AMSB model is specified by the following four parameters :
$\tan\beta, ~{\rm sign}(\mu), ~m_0,~ M_{aux}$. %, 
We scan these parameters over the following ranges :
$ 20~{\rm TeV} \leq  m_{\rm aux} \leq 100~{\rm TeV}$,
$0  \leq  m_0 \leq 2~{\rm TeV}$, $1.5    \leq  \tan\beta \leq 60$, and 
${\rm sign} (\mu)  > 0$. 
In the case of the AMSB scenario with $\mu > 0$, 
the $B\rightarrow X_s \gamma$ constraint is even stronger compared
to other scenarios. 
and  almost all the parameter space with large 
$\tan\beta > 30$ is excluded.
Also stops are relatively heavy in this scenario mainly due to the 
universal addition of $m_0^2$. Therefore the branching ratio for 
$B_s \rightarrow \mu^+ \mu^-$ is smaller than $4 \times 10^{-9}$, 
and this process becomes unobservable at the Tevatron Run II. 
For the detailed discussions on other variations of AMSB scenarios, 
see Refs.~\cite{baek_ko_song}.

Summarizing this section, we showed that there are qualitative differences 
in correlations among $(g-2)_{\mu}$, $B\rightarrow X_s \gamma$, 
and $B_s \rightarrow \mu^+ \mu^-$ in various models for SUSY breaking 
mediation mechanisms, even if all of them can accommodate the muon $a_\mu$:
$10\times 10^{-10} \lesssim a_\mu^{\rm SUSY} \lesssim 40 \times 10^{-10}$. 
Especially, if the $B_s \rightarrow \mu^+ \mu^-$ decay is observed at the 
Tevatron Run II with the branching ratio greater than $2 \times 10^{-8}$, 
the GMSB with low number of messenger fields $N$ and 
certain class of AMSB scenarios would be excluded. 
On the other hand, the minimal supergravity scenario and similar mechanisms 
derived from string models, GMSB with large messenger scale and the 
deflected AMSB scenario can accommodate this observation 
without difficulty for large $\tan\beta$ \cite{baek_ko_song}. 
Therefore search for $B_s \rightarrow \mu^+ \mu^-$ decay at the Tevatron 
Run II would provide us with important informations on the SUSY breaking 
mediation mechanisms, independent of informations from direct search for 
SUSY particles at high energy colliders. 
This is remarkable, since $B_s \rightarrow \mu^+ \mu^-$ could be an excellent
discriminator of SUSY breaking mediations without directly producing SUSY 
particles at all. Let us stay tuned with updated data analysis on this decay
by CDF and D0 collaborations at the Tevatron. 

%%%%%%%%%%%%%%%%%%%%%%%%%%%%%%%%%%%%%%%%%%%%%%%%%%%%%%%%%%%%%%%%%%%%%%%%%%
\section{Interplay of B physics  with cosmology}
\subsection{B physics and electroweak baryogenesis (EWBGEN) 
within SUSY models}

Let us first discuss 
an effective SUSY model with minimal flavor violation \cite{decoupling}. 
In this model, the 1st and the 2nd generation squarks are very heavy and 
almost degenerate, thus evading SUSY flavor/CP problem.
And flavor violation comes through CKM matrix, whereas
CP violation originates from the $\mu$ and $A_t$ phases as well as the KM 
phase. Therefore the stop-chargino loop have additioncal source of CP 
violation in addition to the KM phase in the SM.
One-loop electric dipole moment (EDM) constraint is evaded in the
effective SUSY model due to the decoupling of the 1st/2nd generation 
sfermions, but there are poentially large two-loop contribution
to electron/neutron EDM's through Barr-Zee type diagram in the large 
$\tan\beta$ region \cite{Chang:1998uc}. 
Imposing this two-loop EDM constraint and direct search
limits on Higgs and SUSY particles, we make the following observations
\cite{baek1,baek2}: 
\begin{itemize}
\item No new phase shifts in $B_d - \overline{B}_d$ and 
$B_s - \overline{B}_s$ mixings: 
Time dependent CP asymmetries in $B_d \rightarrow J/\psi K_S$ still 
measures the KM angle $\beta = \phi_1$ 
\item $\Delta M_{B_d}$ can be enhanced upto $\sim 80 \%$ compared to the 
SM prediction 
\item Direct CP asymmetry in $B\rightarrow X_s \gamma$ 
($A_{\rm CP}^{b\rightarrow s \gamma}$) can be as large as $\pm 15 \%$
( Fig.~8 (a) and (b) )   
which is now strongly constrained by the data $(0.5 \pm 3.6) \%$ \cite{acp}
\item $R_{\mu\mu} \equiv B( B\rightarrow X_s \mu^+ \mu^-) / 
B( B\rightarrow X_s \mu^+ \mu^-)_{\rm SM} $ can be as large as 1.8, which 
is now strongly constrained by the data from B factories \cite{work} 
\item $\epsilon_K$ can differ from the SM value by $\sim 40 \%$ \ . 
\end{itemize}
\begin{figure}
\begin{center}
\includegraphics[height=5.5cm]{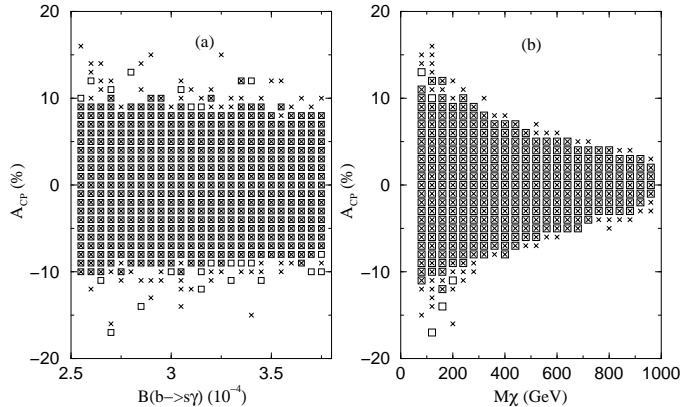}
\end{center}
\caption{Direct CP asymmetry in $B\rightarrow X_s \gamma$ as functions of
(a) $B ( b \rightarrow s \gamma)$ and (b) $m_{\chi}$ (the lighter chargino
mass). The parameter space excluded by two loop EDM constraint is denoted 
by x.}
\end{figure} 
Therefore we predict  substantial deviations in certain observables in the 
$B$ and $K$ systems in SUSY models with minimal flavor violation and complex
$\mu$ and $A_t$ parameters. 
Even if the $A_t$ phase is set to zero, the predictoins do not change much. 
Now this model is beginning to be strongly constrained by
new data on the direct CP asymmetry in $B\rightarrow X_s \gamma$ 
and $R_{\mu\mu}$ from B factories \cite{work}. 

This class of models includes the electroweak baryogenesis (EWBGEN) within 
the MSSM \cite{ewbgen} and some of its extensions such as NMSSM or 
extra $U(1)$ gauge symmetry, where the chargino and stop sectors are the 
same as in the MSSM and the $\mu$ phase plays a key role to generate 
baryon number asymmetry.  
In the EWBGEN scenario within MSSM, Murayama and Pierce argued that there
could no large CP violating effects from the $\mu$ phase on $B$ physics, 
except for the $B_{d(s)} - \overline{B_{d(s)}}$ 
mixing: $1 \leq \Delta M_s /  ( \Delta M_s )_{\rm SM} \leq 1.30$ 
\cite{murayama}.
This is mainly because of the strong tension between the light 
$\tilde{t}_R$ and heavy $\tilde{t}_L$. In the EWBGEN scenario, we  
need a strong 1st order phase transition, and this requires 
a light $\tilde{t}_R$.  On the other hand,  the current LEP bound on the 
lightest Higgs mass $m_h^0$ for  $3 \lesssim \tan\beta \lesssim 6$  
(for larger $\tan\beta$, the $\mu$ phase effect drops out) 
calls for heavy $\tilde{t}_L$ to generate large stop loop corrections to 
$m_h^0$. 
 
However, the LEP bound on the lightest Higgs mass becomes less problematic 
in the extensions of MSSM such as NMSSM or MSSM with extra $U(1)$ 
gauge group, because there are tree level contributions to the Higgs mass.
Therefore the tension between the light $\tilde{t}_R$ and the heavy 
$\tilde{t}_L$ becomes much milder compared to the MSSM, and 
our predictions on the B system  still remain valid in such scenarios.   

%------------------------------------------------------------------------

\subsection{Neutralino dark matter scattering and 
$B_s \rightarrow \mu^+ \mu^-$}

In SUSY models with $R-$parity conservation, the lightest superparticle
(LSP) is stable and becomes a good candidate for dark matter of the 
universe. In particular, the neutralino ($\chi$) LSP is a nice candidate
for cold dark matter, and could be detected in the laboratory through 
(in)elastic scattering with nuclei. There are several direct search 
experiments going on around the world  \cite{Munoz:2003wx}. 
A few years ago, the DAMA Collaboration reported 
a positive signal in the range of $\sigma_{\chi p} = (10^{-5} - 10^{-6} )$ 
pb with $m_{\chi}$ at electroweak scale  \cite{Belli:2002yt}. 
However this was not confirmed by other experiments.
Recently, the DAMA signal region has been excluded
by the CDMS cryogenic  DM search experiment \cite{cdms} in the range of
\[
\sigma_{\tilde{\chi} p} = (10^{-6} - 10^{-5} )~{\rm pb},  
\]
with the corresponding DM mass depends on galactic halo models.
Since the CDMS experiment probes the DM scattering down to
$3 \times 10^{-7}$ pb level in certain range of DM mass,
it is important to calculate the DM scattering cross section within well
defined and/or motivated SUSY models, in which the cross section can be
in the CDMS sensitivity.
Anyway the present sensitivity of the ongoing  
DM scattering experiments is roughly $10^{-6}$ pb, and it is important to
identify the parameter space of general MSSM which can be probed by the DM
scattering experiments. 

In the following we show that there is a strong correlation between
$\sigma_{\chi p}$ and $B_s \rightarrow \mu^+ \mu^-$ \cite{Baek:2004et}. 
In the large $\tan\beta$ region of SUSY models, both processes are 
dominated by neutral Higgs exchange diagram,  and the amplitudes 
for these two processes  depend on $\tan\beta$ as 
\begin{eqnarray}
{\cal M} ( B_s \rightarrow \mu^+ \mu^- ) & \propto & \tan^3 \beta / m_A^2 ,
\nonumber  \\
{\cal M} ( \chi^0 p \rightarrow \chi^0 p ) & 
\propto & \tan\beta / m_A^2 .
\end{eqnarray}
Therefore one can expect some correlation between the two obervables
in the large $\tan \beta$ limit. Since the current limit on 
$B ( B_s \rightarrow \mu^+ \mu^- )$ is already tight enough, this could 
provide an important constraint on the neutralino DM scattering cross 
section.   

In the minimal supergravity model with $R-$parity conservation,
the LSP is binolike neutralino in most parameter space, and the 
spin-independent dark matter scattering cross section $\sigma_{\chi p}$
turns out to be very small $\lesssim 10^{-8}$ pb, after imposing various
constraints from Higgs and SUSY particle masses, $B\rightarrow X_s \gamma$,
etc. [ Fig.~9 (a) ]. 
The  mSUGRA models cannot give a large enough $\sigma_{\chi p}$ 
in the signal regions of DAMA and CDMS or in the sensitivity region 
of other experiments down to $\sim 10^{-8}$ pb.
However, the usual minimal SUGRA boundary conditions for soft 
parameters are too much restrictive without theoretical justification, 
and it is important to study the dark matter scattering in  more general 
supergravity models with nonuniversal soft terms
\cite{cerdeno}. In such case, one has to be careful not to overproduce 
flavor changing neutral current processes, which is a subject of this 
subsection.

\begin{figure*}
\begin{tabular}{ccc}
{\includegraphics[height=5.0cm]{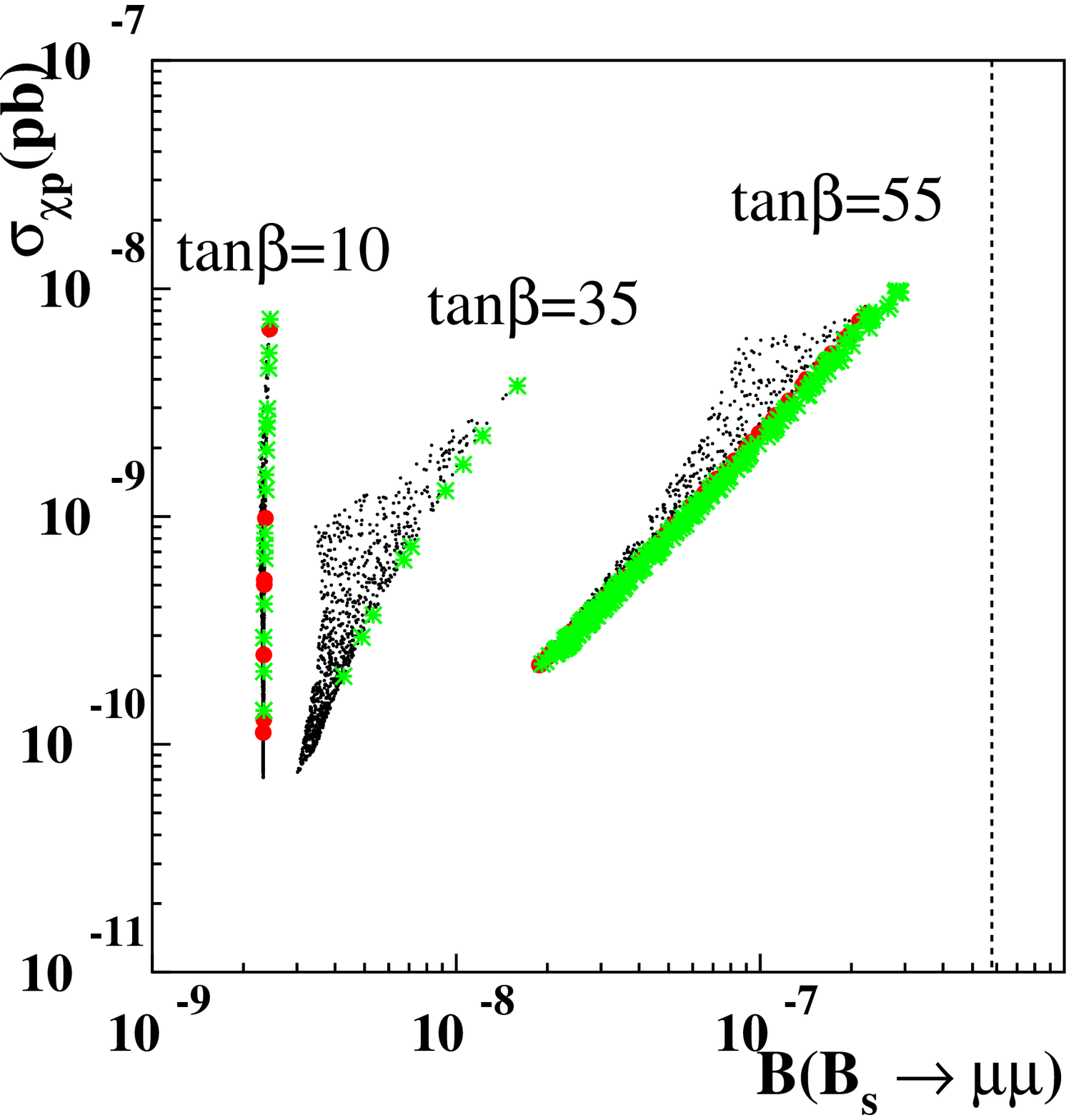}}&
{\includegraphics[height=5.0cm]{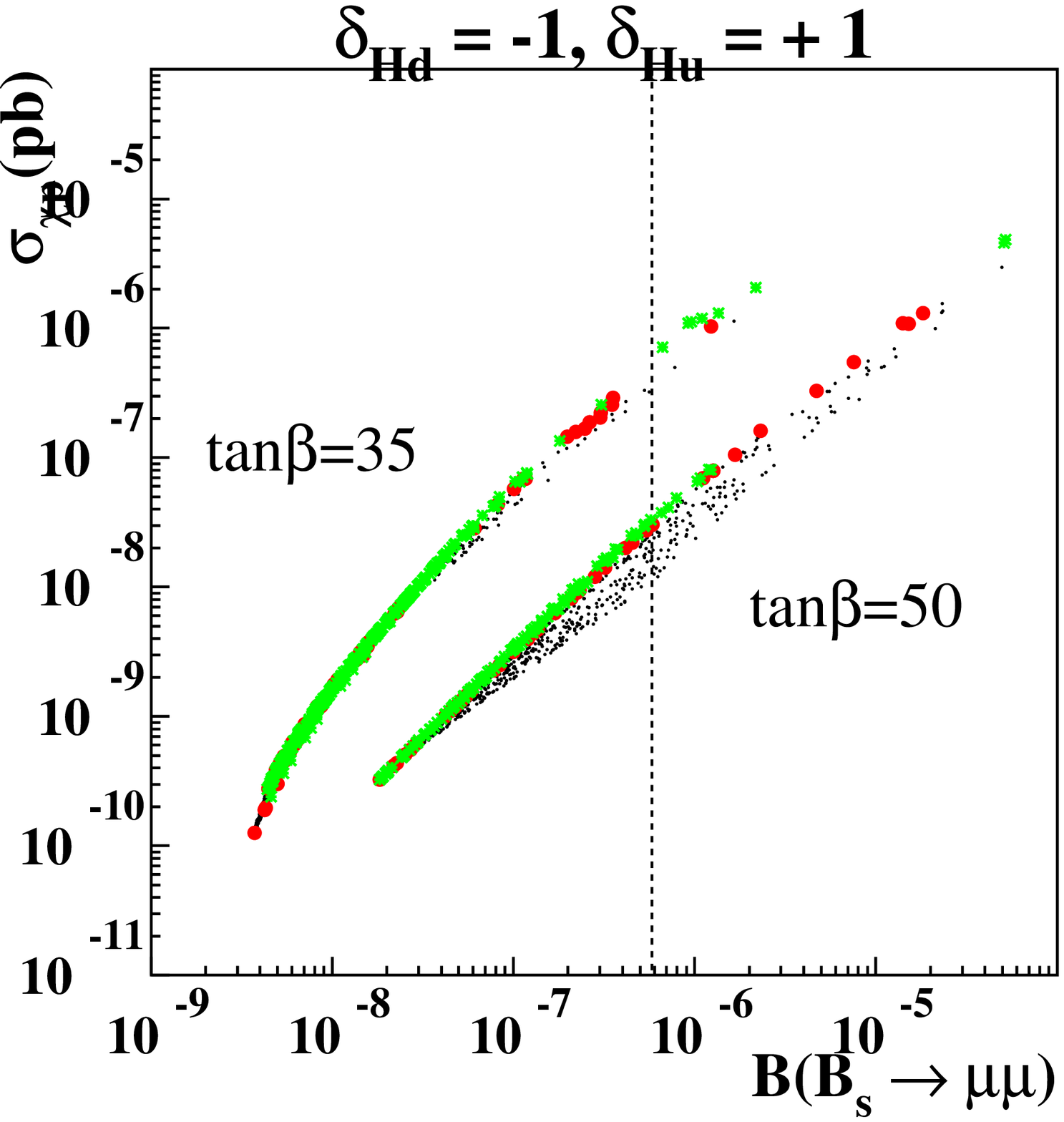}}&
{\includegraphics[height=5.0cm]{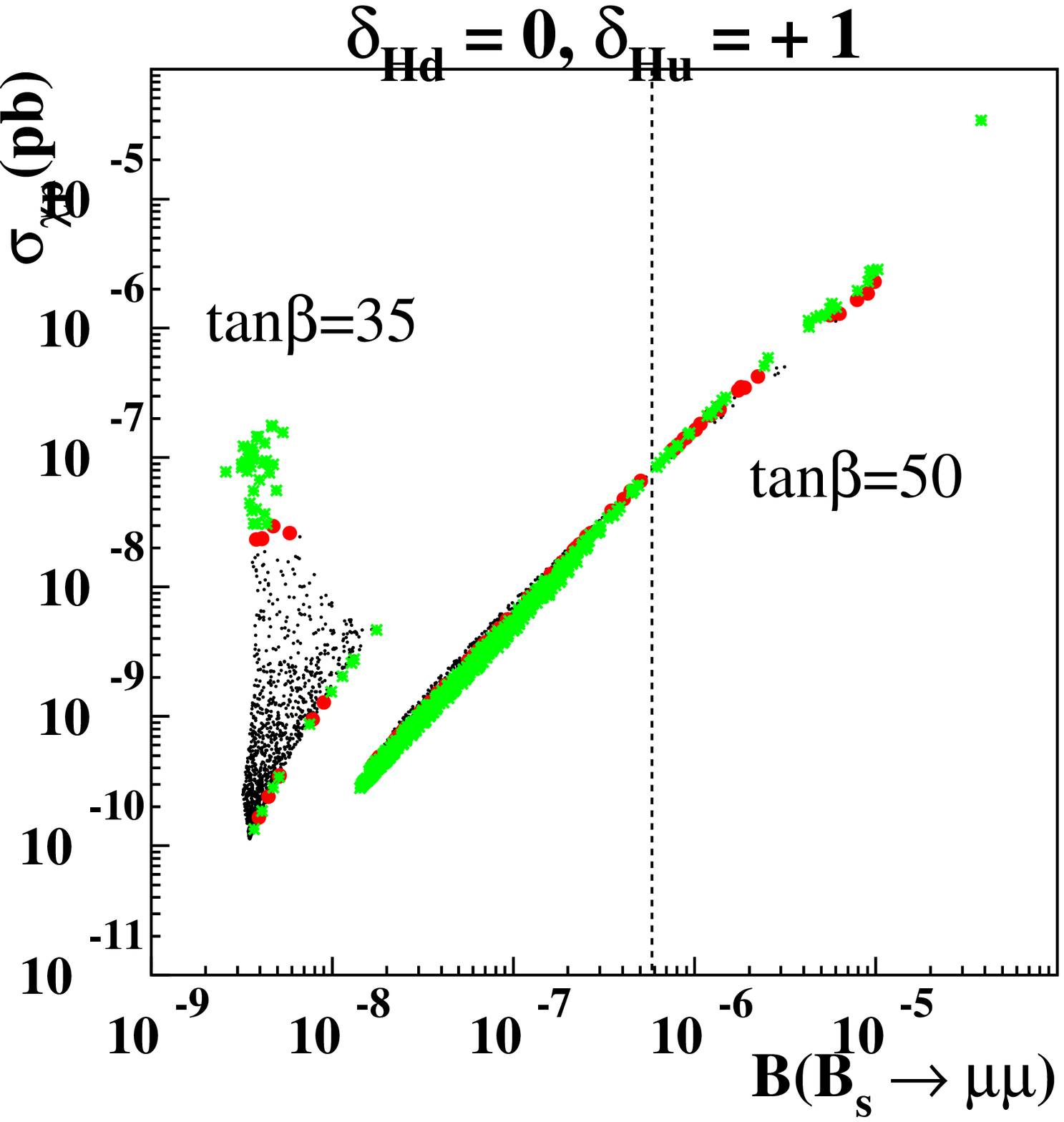}}
\end{tabular}
\caption{\label{fig3} 
$\sigma_{\tilde{\chi} p}$ vs. $B( B_s \rightarrow \mu^+ \mu^- )$ within 
(a) mSUGRA with universal Higgs mass parameters for $\tan\beta = 10, 35 $ 
and 55 (from the left to the right),  
in SUGRA with nonuniversal Higgs mass parameters: 
(b) $\delta_{H_u} = 1$ and $\delta_{H_d} = -1$ and 
(c) $\delta_{H_u} = 1$ and $\delta_{H_d} = 0$. 
Black dots for $\Omega_\chi h^2 \geq 0.13$, 
red dots for $0.095 \leq \Omega_\chi h^2 \leq 0.13$ and
green dots for $\Omega_\chi h^2 \leq 0.095$.
}
\end{figure*}

As discussed before, the universal soft parameters are too restricted 
assumption without solid ground within supergravity framework. In order 
to consider more generic situation within supergravity scenario, let us 
relax the assumption of  universal soft masses as follows: 
\begin{equation}
m_{H_u}^2 = m_0^2 ~( 1 + \delta_{H_u} ),~~~
m_{H_d}^2 = m_0^2 ~( 1 + \delta_{H_d} ),~~~
\end{equation}
whereas other scalar masses are still universal.  Here $\delta$'s are 
parameters with $\lesssim O(1)$. By allowing nonuniversality in the 
Higgs mass parameters, the situation changes, however.  For illustration 
of our main point, let us  take the numerical values of $\delta$'s as in 
Refs.~\cite{Munoz:2003wx,cerdeno}: 
\begin{eqnarray}
(I) &  \delta_{H_d} = -1 , &  \delta_{H_u} = 1 ,
\nonumber  \\
(II) &  \delta_{H_d} = 0 , &  \delta_{H_u} = 1 ,
\end{eqnarray}
For $\delta_{H_u} = +1$, $\mu$ becomes lower and the Higgsino component
in the neutralino LSP increases so that $\sigma_{\chi p}$ is enhanced, as
discussed in Ref.~\cite{Munoz:2003wx}. 
The change of $|\mu|$ also has an impact on the higgs masses because
\[
m_A^2 = m_{H_u}^2 + m_{H_d}^2 + 2\mu^2 \simeq m_{H_d}^2 + \mu^2 - M_Z^2/2
\] 
at weak scale.  For $\delta_{H_d} = -1$, $m_A$ and 
$m_H$ becomes  further  lower,  and both $\sigma_{\chi p}$ and 
$B( B_s \rightarrow \mu^+ \mu^- )$ are enhanced. These features are 
shown in Fig.s~\ref{fig3} (b) and (c) for Case (I) and (II), respectively. 
Note that the CDF  upper bound 
$B ( B_s \rightarrow \mu^+ \mu^- ) < 5.8 \times 10^{-7}$ \cite{cdf} 
provides a very strong constraint on the neutralino DM scattering 
cross section $\sigma_{\chi p}$, and removes the parameter space where 
the DM scattering is within the reach of the current DM search experiments. 

The impact of the $B ( B_s \rightarrow \mu^+ \mu^- )$ branching ratio
constraint becomes more transparent in Fig.s~10 (a) and (b). 
Here we plot the neutralino DM scattering cross sections as functions 
of the LSP mass, imposing the CDMS bound as well as the 
$B ( B_s \rightarrow \mu^+ \mu^- )$. The red points in the plots are 
excluded by $B ( B_s \rightarrow \mu^+ \mu^- ) < 5.7 \times 10^{-7}$.   
For $\tan\beta = 35$, both constraints are comparable. 
However,  for $\tan\beta = 50$, the $B_s \rightarrow \mu^+ \mu^- $ 
branching ratio puts a much stronger  constraint than the direct search
by CDMS. This clearly shows the importance of $B_s \rightarrow \mu^+ \mu^-$
when we study the neutralino DM scattering.

\begin{figure*}
%\begin{tabular}{cc}
\begin{center}
{\includegraphics[height=5.0cm]{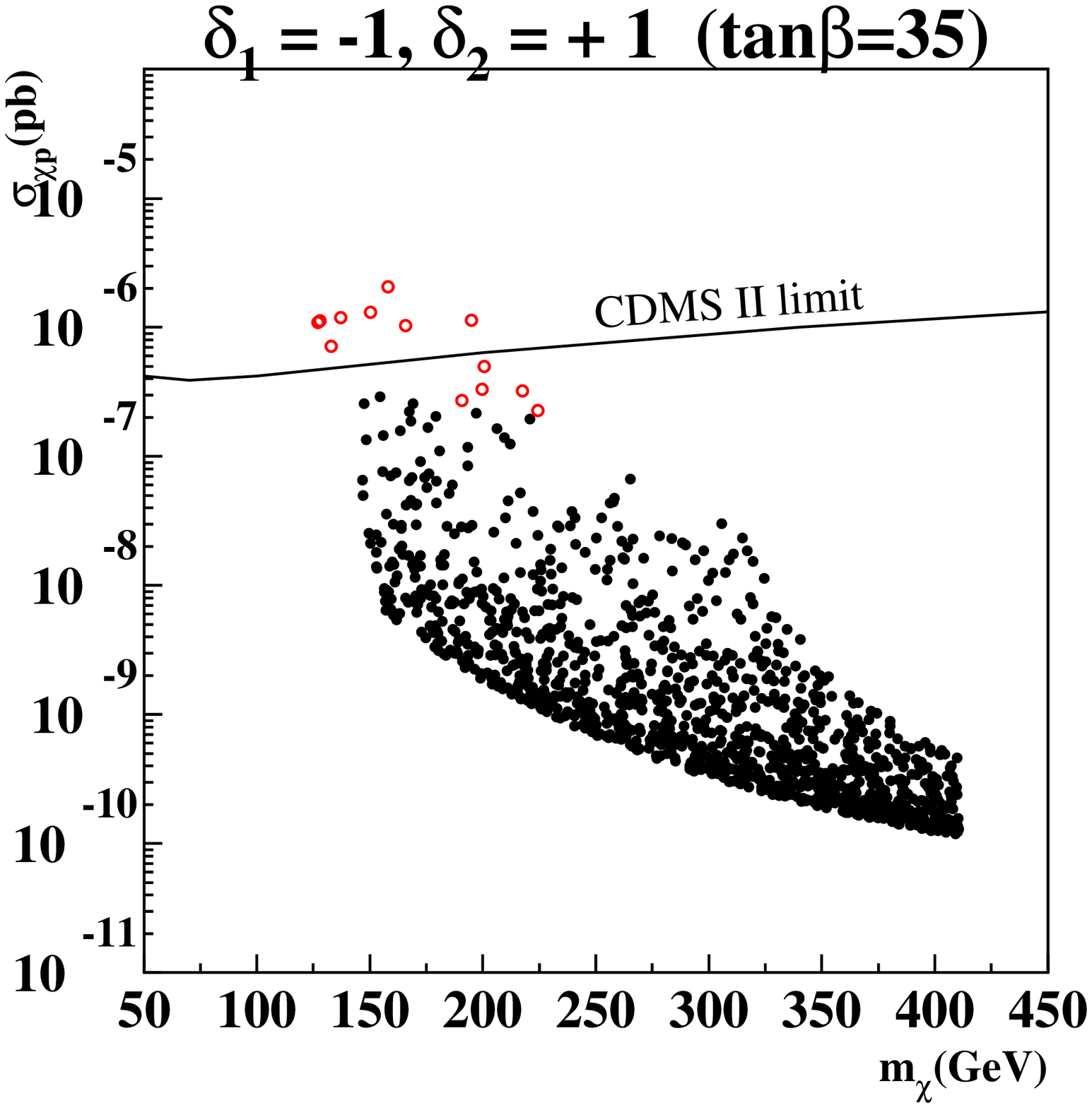}} %&
%{\includegraphics[height=5.0cm]{nhiggsm11.eps}}&
{\includegraphics[height=5.0cm]{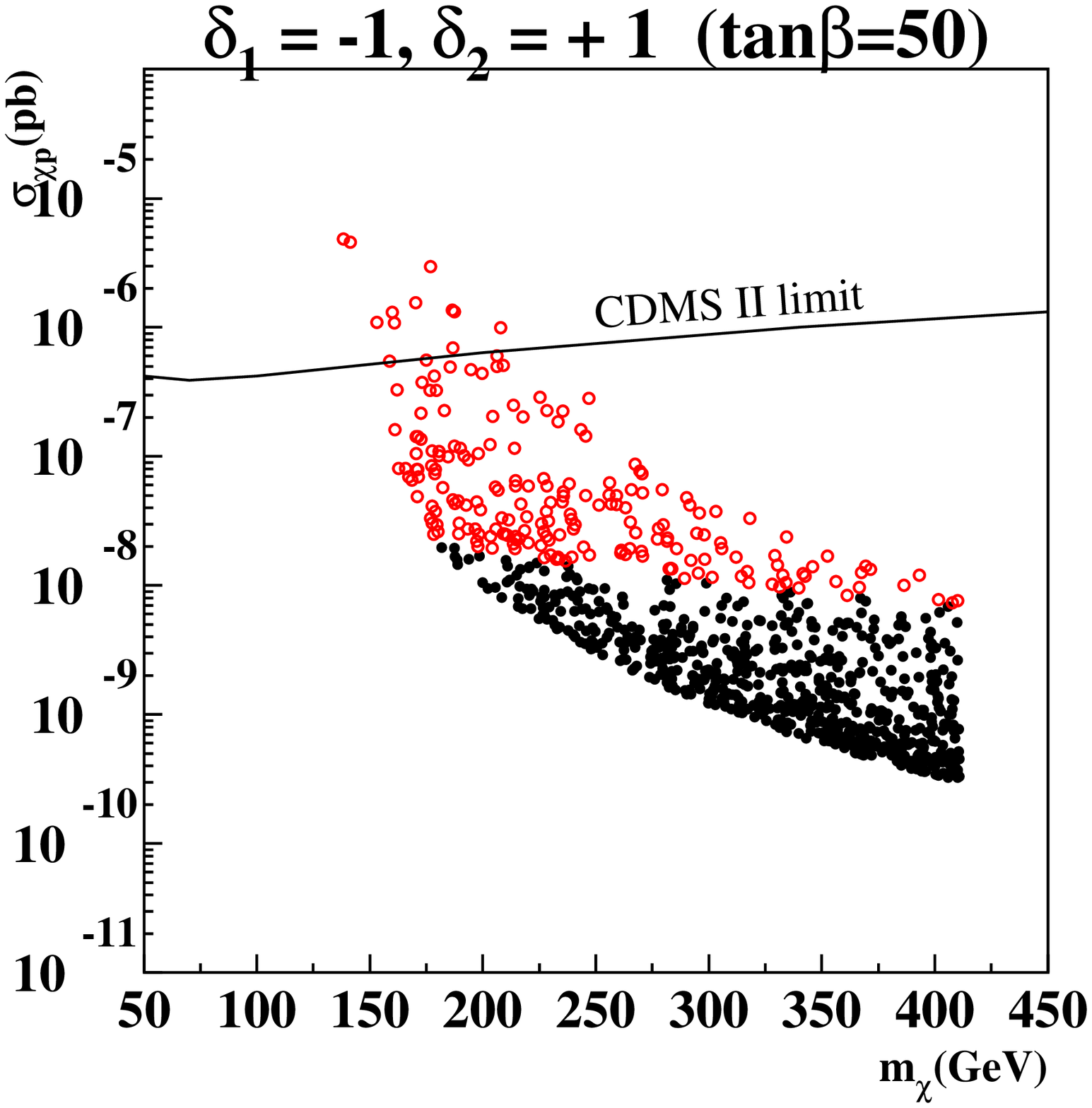}}
%\end{tabular}
\end{center}
\caption{\label{fig4}
$\sigma_{\tilde{\chi} p}$ vs. the neutralino LSP mass, with the CDMS 
upper bound. The points which are excluded by 
$B( B_s \rightarrow \mu^+ \mu^- ) < 5.7 \times 10^{-7}$ 
are denoted by red dots. 
}
\end{figure*}

We also considered nonuniversal gaugino masses, in which case the most 
important one is the gluino mass parameter via RG running. 
Therefore we allowed nonuniversality only in the gluino mass parameter, and
found that the qualitative  feature is similar 
as in nonuniversal Higgs masses.  In particular the current limit on 
$B ( B_s \rightarrow  \mu^+ \mu^- )$ already puts a strong constraint on 
$\sigma_{\tilde{\chi} p}$ in the large $\tan\beta$ region.

In summary, we found that the upper limit on 
$B ( B_s \rightarrow \mu^+ \mu^- )$ is an important constraint on SUSY 
parameter space in the large $\tan\beta$ region, and the DM scattering 
cross section could be strongly affected by this constraint. This is an 
example of an interesting interplay between particle physics and cosmology.

%%%%%%%%%%%%%%%%%%%%%%%%%%%%%%%%%%%%%%%%%%%%%%%%%%%%%%%%%%%%%%%%%%%%%%%%%%
\section{Conclusion}

In this talk, I discussed flavor physics within SUSY models, 
in particular where we may expect large deviations from the SM predictions, 
even if the unitarity triangle is the same as the SM case. This includes 
$B\rightarrow X_d \gamma$, $B\rightarrow X_s \gamma$,
$B_d \rightarrow \phi K_S$, $B_s - \overline{B}_s$ mixing, and 
$B_s \rightarrow \mu^+ \mu^-$ as well as $\epsilon^{'} / \epsilon_K$. 
Also I discussed some interplay between B physics and cosmologically 
interesting SUSY scenarios.  In  EWBGEN scenarios within SUSY models, 
one may expect a large direct CP violation in $B\rightarrow X_s \gamma$, 
which is now strongly constrained by the data.  Dark matter scattering 
cross section and $B_s \rightarrow \mu^+ \mu^-$ exhibit a strong 
correlation for large $\tan\beta$. In particular, the branching ratio 
of $B_s \rightarrow \mu^+ \mu^- $ can exceed the current CDF limit, 
when the DM scattering cross section becomes large within the sensitivity 
of the current DM search experiments: $\sigma_{\chi p} \sim ( 10^{-6} - 
10^{-7} )$ pb. 
This is an example where B physics and cosmology show an interesting 
interplay, and the upper limit on the branching ratio for 
$B_s \rightarrow \mu^+ \mu^-$ becomes an important constraint on SUSY 
parameter space in the large $\tan\beta$ region, even stronger than the
CDMS bound.   
In short, it is still possible to have  substantial SUSY effects in the 
$b \rightarrow s$ transition without conflict with any other observed 
phenomena as of now. Therefore these processes should be actively searched 
for at B factory experiments in the coming years. By doing so, we can 
verify  the CKM paradigm for flavor and CP violation, 
and better constrain the flavor and CP structures of SUSY models. 
Or we may encounter some nice surprise from the $b\rightarrow s$ transition.

Note : Recently, the D0 collaboration presented a new data \cite{d0}:
\[
B ( B_s \rightarrow \mu^+ \mu^- ) < 4 \times 10^{-7} ~~~
(90 \% ~{\rm C.L.}),
\]
which is better than the CDF bound we used in the published paper. 
This D0 data would improve slightly the bounds discussed in Sec.~4 
and Sec.~5.2. 

\section{Acknowledgements}

I thank the orgranizer of the 1st GUC workshop Prof. Shabaan Khalil 
for inviting me to this meeting.  
I am also grateful to my collaborators, Seungwon Baek, Gordy L. Kane, 
Yeong-Gyun Kim, Christopher Kolda, Gustav Kramer, Antonio Masiero, 
Jae-hyeon Park, Wan Young Song, Haibin Wang and Lian-tao Wang, 
for enjoyable collaborations on  the works presented in this talk. 
This work is supported in part by the BK21 Haeksim Program, 
KRF grant KRF-2002-070-C00022 and KOSEF through CHEP at Kyungpook National 
University, and by KOSEF Sundo Grant R02-2003-000-10085-0. 

\bibliographystyle{plain}

%\noindent  
%It is preferred that references in the bibliography be cited in the text 
%using an Arabic number in a square bracket; 
%for example, if we cite the paper by Cohen and Anderson the reference 
%number appears like this~\cite{paper2}.
%All references should include initials and last name of the author(s), 
%volume (in bold), year of publication of paper in the journal/book, 
%and page/article numbers.  
%The journal name and the title of publication should be in italics 
%as in Refs.~\cite{paper1,paper2,paper3}.   

\end{document}